\documentclass{aa}  

\newcommand{\Halpha}{H$\alpha$}
\newcommand{\HII}{H\textsc{ii}}
\newcommand{\Htwo}{H$_2$}

\usepackage{graphicx}
\usepackage[dvipsnames]{xcolor}
\usepackage{txfonts}
\usepackage{multirow}
\usepackage[colorlinks=true,citecolor=blue,urlcolor=blue,linkcolor=blue]{hyperref}
\begin{document}

\title{The impact of spiral arms on the star formation life cycle}

\author{Andrea~Romanelli\inst{1}\and Mélanie~Chevance\inst{1,2} \and J.~M.~Diederik~Kruijssen\inst{3,2} \and Lise~Ramambason\inst{1} \and Miguel~Querejeta\inst{4} \and
Mederic~Boquien\inst{5} \and
Daniel~A.~Dale\inst{6} \and
Jakob~den~Brok\inst{7} \and
Simon~C.~O.~Glover\inst{1} \and
Kathryn~Grasha\inst{8} \and
Annie Hughes\inst{9} \and
Jaeyeon Kim\inst{10}\and
Steven Longmore\inst{11,2} \and
Sharon E. Meidt\inst{12} \and 
Jos\'e Eduardo Mendez-Delgado\inst{13,14}\and
Lukas~Neumann\inst{15}\and
Jérôme Pety\inst{16,17}
Eva Schinnerer\inst{18}
Rowan Smith\inst{19}\and
Jiayi Sun\inst{20}\and
Thomas G.~Williams\inst{21}
}

\institute{
Zentrum für Astronomie der Universit\"{a}t Heidelberg, Institut für Theoretische Astrophysik, Albert-Ueberle-Str. 2, 69120 Heidelberg \\
email: \href{mailto:andrea.romanelli@stud.uni-heidelberg.de}{andrea.romanelli@stud.uni-heidelberg.de}
\and 
Cosmic Origins Of Life (COOL) Research DAO, \href{https://coolresearch.io}{https://coolresearch.io}
\and 
Technical University of Munich, School of Engineering and Design, Department of Aerospace and Geodesy, Chair of Remote Sensing Technology, Arcisstr.~21, 80333 Munich, Germany
\and 
Observatorio Astronómico Nacional (IGN), C/Alfonso XII, 3, E-28014 Madrid, Spain
\and 
Université Côte d’Azur, Observatoire de la Côte d’Azur, CNRS, Laboratoire Lagrange, 06000, Nice, France
\and 
Department of Physics and Astronomy, University of Wyoming, Laramie, WY 82071, USA
\and 
Center for Astrophysics | Harvard \& Smithsonian, 60 Garden Street, 02138 Cambridge, MA, USA
\and 
Research School of Astronomy and Astrophysics, Australian National University, Canberra, ACT 2611, Australia
\and IRAP, OMP, Université de Toulouse, 9 Av. du Colonel Roche, BP 44346, F-31028 Toulouse cedex 4, France
\and
Kavli Institute for Particle Astrophysics \& Cosmology, Stanford University, CA 94305, USA
\and
Astrophysics Research Institute, Liverpool John Moores University, IC2, Liverpool Science Park, 146 Brownlow Hill, Liverpool L3 5RF, UK
\and 
Sterrenkundig Observatorium, Universiteit Gent, Krijgslaan 281 S9, B-9000 Gent, Belgium
\and 
Astronomisches Rechen-Institut, Zentrum für Astronomie der Universität Heidelberg, Mönchhofstr. 12-14, D-69120 Heidelberg, Germany
\and
Instituto de Astronomía, Universidad Nacional Autónoma de México, A.P. 70-264, 04510 México D. F., México
\and
European Southern Observatory, Karl-Schwarzschild Stra{\ss}e 2, D-85748 Garching bei M\"{u}nchen, Germany
\and
IRAM, 300 rue de la Piscine, 38400 Saint Martin d'H\`eres, France
\and
LUX, Observatoire de Paris, PSL Research University, CNRS, Sorbonne Universités, 75014 Paris, France
\and
Max Planck Institute for Astronomy, Königstuhl 17, 69117 Heidelberg, Germany 
\and
SUPA, School of Physics and Astronomy, University of St Andrews, North Haugh, St Andrews, KY16 9SS, UK
\and Department of Astrophysical Sciences, Princeton University, 4 Ivy Lane, Princeton, NJ 08544, USA
\and Sub-department of Astrophysics, Department of Physics, University of Oxford, Keble Road, Oxford OX1 3RH, UK
}

\date{Received XXX / Accepted XXX}

\abstract{
The matter cycle between gas clouds and stars in galaxies plays a crucial role in regulating galaxy evolution through feedback mechanisms. In turn, the local and global galactic environments shape the interstellar medium and provide the initial conditions for star formation, potentially affecting the properties of this small-scale matter cycle. In particular, spiral arms have been proposed to play a pivotal role in the star formation life cycle, by enhancing the gas density and triggering star formation. However, their exact role is still debated. In this study, we investigated the role of spiral arms in the giant molecular cloud evolutionary life cycle and on the star formation process in a sample of 22 nearby spiral galaxies from the PHANGS survey. We measured the cloud lifetime, the feedback timescale, the typical distance between independent regions, and the star formation efficiency in spiral arms and inter-arm regions separately. We find that the distributions of the cloud lifetime as well as the feedback timescale are similar in both environments. This result suggests that spiral arms are unlikely to play a dominant role in triggering star formation. By contrast, the star formation efficiency appears to be slightly higher in inter-arm regions compared to spiral arms.
}


\titlerunning{The impact of spiral arms on the star formation life cycle}
\authorrunning{A.\ Romanelli et al.}

\maketitle
\section{Introduction}

Galaxies in the universe show a variety of morphological structures, which play a pivotal role in galactic evolution due to the strong link between the local galactic environment and the ability of gas to form new stars. Examples of this relation are the accumulation of molecular gas in spiral structures and rings \citep{Dobbs2011} and the morphological quenching of star formation in galaxy spheroids \citep{Martig2009,davis14,gensior20}. In particular, star formation in the local universe has been shown to preferentially occur in high-surface brightness spiral galaxies \citep{Brinchmann04}. In this respect, spiral structures have been proposed to enhance star formation rates by collecting gas from the diffuse galactic medium that falls into its potential well, thereby increasing the local gas density and, therefore, creating favourable conditions for star formation \citep{roberts69,dobbs14}. However, it has long been debated whether galactic dynamics has a triggering effect on the star formation process itself \citep{elmegreen86}, and our understanding of how galactic morphological structures affect the conversion of molecular gas into stars remains insufficient \citep{chevance23,schinnerer&leroy24}.

Recent studies show that the properties (e.g.\ mass, radius, and average velocity dispersion) of giant molecular clouds (GMCs), which are the pivotal sites for star formation, vary among and within galaxies, as a function of the local (dynamical) environment \citep[e.g.][]{hughes13, colombo14, rosolowsky21}. GMCs, which were long considered to be quasi-equilibrium structures \citep{scoville79,Koda2009}, are now viewed through a different lens that acknowledges their connection to the galactic potential \citep{meidt18,sun20} and their transient nature shaped by feedback from young massive stars. These young stars replenish the Inter-stellar Medium (ISM) of material, which will form new generations of stars and, simultaneously, inject energy and momentum, which can contribute to globally halting cloud formation. 

The processes regulating the global star formation act on a wide range of scales, from small to large. High-resolution observations have shown that the separation between GMCs and \HII\ regions \citep{schruba10} can be interpreted as a sign of a rapid evolutionary cycling between gas clouds and young stars, as well as stellar feedback acting to destroy the surrounding molecular gas \citep{kruijssen14}. These observations give insights into the action of galactic dynamics at the cloud scale. For example, in M51 the ratio between star formation rate (SFR) and molecular gas mass (i.e.\ the star formation efficiency, SFE) is much lower in the inner parts of the southern arm \citep{meidt13,schinnerer13,leroy17,querejeta19} than elsewhere along the arms. Variations in SFE of this kind have been attributed to a dynamical suppression of star formation \citep{meidt13}.

However, it is still debated if large-scale dynamical features can directly regulate the creation and destruction of molecular clouds. In this respect, spiral shocks are thought to influence the star formation process in two possible ways. On the one hand, they are a means of gathering gas and, in doing so, enhancing the gas mass surface density. In this scenario, the SFE should remain unaffected by the passage of the spiral density waves. On the other hand, spiral shocks also modify the dynamical state of gas clouds (on the ${\sim}100$\,pc scale), potentially triggering their collapse \citep[see e.g.\ the case of star formation induced by GMC-GMC collisions]{fukui14}. Turbulence-regulated star formation models \citep[e.g.][]{krumholz05} predict a larger width of the gas density probability distribution function in environments where the velocity dispersion is larger. This seems to be the case for spiral structures, which suggests an increase amount of gas at high densities, and an enhanced efficiency of star formation. Alternatively, the differential gas flows and shear in spiral arms can provide support against collapse \citep{meidt18}. Tidal forces from nearby clouds can also tear clouds apart in spiral arms. These effects could in turn limit the ability of clouds to form stars. In addition, spiral features are the preferential sites of ongoing high-mass star formation and, therefore, the location where feedback processes by massive stars continuously inject momentum \citep[][and references therein]{Meidt21}. To understand the importance of spiral arms for star formation, we therefore need to determine how the duration of the star formation cycle (i.e.\ the lifetime of GMCs and the duration over which they form stars) and the efficiency of the conversion of gas to stars may differ between spiral arms and inter-arm regions.

Previous studies \citep{Scoville2004,Koda2009,Koda2013} have suggested a predominant role of spiral arms in the star formation life cycle, where GMCs only form in spiral arms, and the presence of GMCs in the inter-arm regions is explained by a long cloud lifetime (${>}100$\,Myr). However, short cloud lifetimes have now been measured with a variety of different methods in nearby galaxies \citep{Blitz1990,Engargiola2003,kawamura09,Meidt2015,kruijssen19,grasha19, chevance20,kim22}, and are found to be comparable with the GMC internal dynamical timescale estimates (${\sim}10{-}50$\,Myr). These results point to a more limited role of the spiral arms in the evolution of GMCs. Different studies also find contradictory results for the environmental dependence of the SFE (${\rm SFE} =\Sigma_{\rm SFR}/\Sigma_{\rm H_{2}}$) or, similarly, the inverse of the depletion timescale ($\tau_{\rm depl} = 1/{\rm SFE}$). While some studies find an enhanced SFE in spiral structures \citep{lord90,knapen96,seigar02,cedres13,ragan18,yu21,chen24}, others do not see any significant difference with respect to the inter-arm regions \citep[e.g.,][]{henry03, foyle10, kreckel16, ragan18, querejeta21,querejeta24,sun&calzetti24}. In addition, the depletion timescale is an instantaneous measure of the current gas consumption rate within (a region of) a galaxy and does not probe the integrated SFE per cloud. Both measurements of the cloud lifetime and of the depletion time are necessary to compare the (integrated) SFE between clouds in spiral arms and inter-arm regions, as well as to assess the role of spiral arms in the star formation life cycle.

In this paper, we leverage the PHANGS survey \citep{Leroy2021a}\footnote{Physics at High Angular resolution in Nearby Galaxies, www.phangs.org} and the environmental masks established by \citet{querejeta21} to identify spiral arms and inter-arm regions to address this question. Before the Atacama Large Millimeter/submillimeter Array (ALMA), many efforts were made to trace the molecular gas at the cloud scale in nearby galaxies \citep{downes98,rosolowsky05, Koda2009, donovan13,schinnerer13}. However, either because of insufficient sample size, field of view, or sensitivity, the systematic study of the link between star formation and galactic structures at high resolution has remained limited. The PHANGS-ALMA survey has now enabled the study of the molecular gas properties and distribution in 90 galaxies (covering a wide range of morphologies), using the CO(2-1) emission line, with a resolution of ${\sim}100$\,pc \citep{Leroy2021a}. These cloud-scale observations of the nearby galaxy population, combined with the statistical method developed by \citet{kruijssen14} and \citet{kruijssen18} (the 'Uncertainty Principle for Star Formation'), enable the systematic characterisation of the life cycle between cloud formation, cloud evolution, and star formation. This statistical method quantifies the cloud-scale variations of the flux ratio between molecular gas and a star formation tracer and translates these into the durations of the visibility of each tracer. Indeed, these small-scale departures from the galactic-scale star formation relation \citep[e.g.\ ][]{Silk97, kennicutt98} stem directly from sampling individual regions independently evolving through the different phases of the same cloud-to-star cycle \citep[e.g.\ ][]{feldmann11, kruijssen14}. As a result, the break-down of the star formation relation at the cloud scale is tightly connected to the duration of each phase of the cycle. We present here the cloud lifetimes, the duration of star formation, and the spatial distribution of star-forming regions in spiral arms and inter-arm regions of a sub-sample of 22 spiral galaxies with a pronounced spiral structure (as identified by \citealt{querejeta21}). Constraining these parameters as a function of local environment enables us to assess the role of spiral arms in the cloud evolution and star formation life cycle.

\section{Data}

\subsection{Sample selection}

The sample used in this paper is built by selecting galaxies for which both CO(2-1) and \Halpha\ observations are available in the PHANGS sample (see \citealt{Leroy2021a} for CO(2-1) and Razza et al.\ in prep.\ for \Halpha). Furthermore, as a requirement for an accurate measurement of the cloud evolutionary timeline using the statistical method developed by \citet{kruijssen18}, a sufficiently large number of emission peaks needs to be detected in each of the two emission maps (see Section~\ref{method}). We therefore select galaxies for which at least 35 peaks with a signal-to-noise ratio greater than 5 are detected in each tracer, for both inter-arm regions and spiral arms. This selection follows the same criteria as in \citet{kim22}, which gathered a sample of 54 galaxies in which the cloud evolutionary cycle is characterised. Out of these 54 galaxies, 24 have an environmental mask available from \citet{querejeta21} with clear spiral structures. We further removed two galaxies (NGC~4731 and NGC~1566) for which the fitting of the cloud life cycle in the individual environments (arms and inter-arm regions) did not converge. In the following, we investigate the environmental dependence of the cloud and star formation cycle in the 22 galaxies listed in Table~\ref{table1}. More details on the properties of the selected galaxies can be found in \citet{kim22}.

\subsection{Molecular gas tracer}

The bright transitions of the CO molecule are commonly used to trace the molecular gas in extragalactic studies \citep[e.g.][]{Kuno07,leroy13, Saintonge17, Sorai19}. Here, we use the public release version of the PHANGS–ALMA CO(2–1) \citep{Leroy2021a} moment-0 maps with the 'broad' masking scheme at native resolution. The maps have been collected combining the extended and compact ALMA arrays (TP+7~m+12~m arrays) and have a resolution of $\sim$1\arcsec, which translates into GMC-scale resolution ($\sim$80-150\,pc) at the distances of the galaxies in our sample and enables an effective analysis of feedback and cloud lifetimes. More information on the CO maps can be found in \citet{Leroy2021a}, with details about the data-reduction pipeline provided in \citet{leroy21_pipe}.

\subsection{Star formation tracer}

The \Halpha\ line originates from the ionised surroundings of a newborn high-mass star, and it is a conspicuous feature of \HII\ regions. As such, \Halpha\ is commonly used as a tracer of the current star formation rate \citep[SFR,][]{kennicutt12}. To trace the SFR, we use the continuum-subtracted narrow-band \Halpha\ imaging from PHANGS–\Halpha\ (Razza et al.\ in prep.). Details about the specific data reduction steps can be found in \citet{schinnerer19} and \citet{kim22}; here we provide only a brief summary. The PHANGS–\Halpha\ sample consists of 65 galaxies, observed with a narrow-band filter using the du Pont 2.5~m telescope at the Las Campanas Observatory and using the Wide Field Imager (WFI) instrument at the MPG-ESO 2.2~m telescope at the La Silla Observatory, with a typical resolution of 1\arcsec. All the galaxies are corrected for the Milky Way dust extinction, using a \citet{fitzpatrick99} extinction curve with $R_V = 3.1$. However, we do not correct for internal, spatially resolved extinction from gas and dust around young stars. \citet{haydon20_ext} verified that embedded HII regions can potentially affect the measurement of the overlap phase, but only for gas surface densities above $20\,\mathrm{M_{\odot}\,pc^{-2}}$ at solar metallicity. Most of the galaxies in our sample fall below this threshold \citep{kim22}. All \Halpha\ maps are background subtracted, corrected for the loss due to the filter transmission, and corrected for the contribution of the [NII] lines at 654.8 and 658.3\,nm to the narrow-band filter flux (see \citealt{schinnerer19} and Razza et al.\ in prep.\ for details).

\subsection{Environmental masks}\label{envmask}

In this paper, we seek to investigate the role of spiral arms in the cloud and star formation life cycle in galaxies. We therefore divide the galaxies in our sample into 'spiral arms' and 'inter-arm regions' based on the environmental masks established by \citet{querejeta21}. These environmental masks were constructed based on Spitzer IRAC $3.6~\mu$m observations, with a resolution of 1\farcs7. The masks are purely morphological and led to the definition of seven distinct environments for a sample of 74 PHANGS-ALMA galaxies: bars, spiral arms, rings, lenses, bulges, centres, and discs. For the purpose of this paper, we are interested in two types of environments: the area that falls within the spiral structures and its counterpart, the inter-arm regions. Because the analysed field of view is limited by the rather restrictive ALMA coverage, both environments span the full range of available galactocentric radii for each galaxy. We note that, as a prerequisite to the application of our statistical method (see Sect.~\ref{method}), we exclude the central parts of all galaxies affected by blending \citep[see e.g. ][]{chevance20}. 

\begin{figure}[tbp]
\includegraphics[width=\columnwidth]{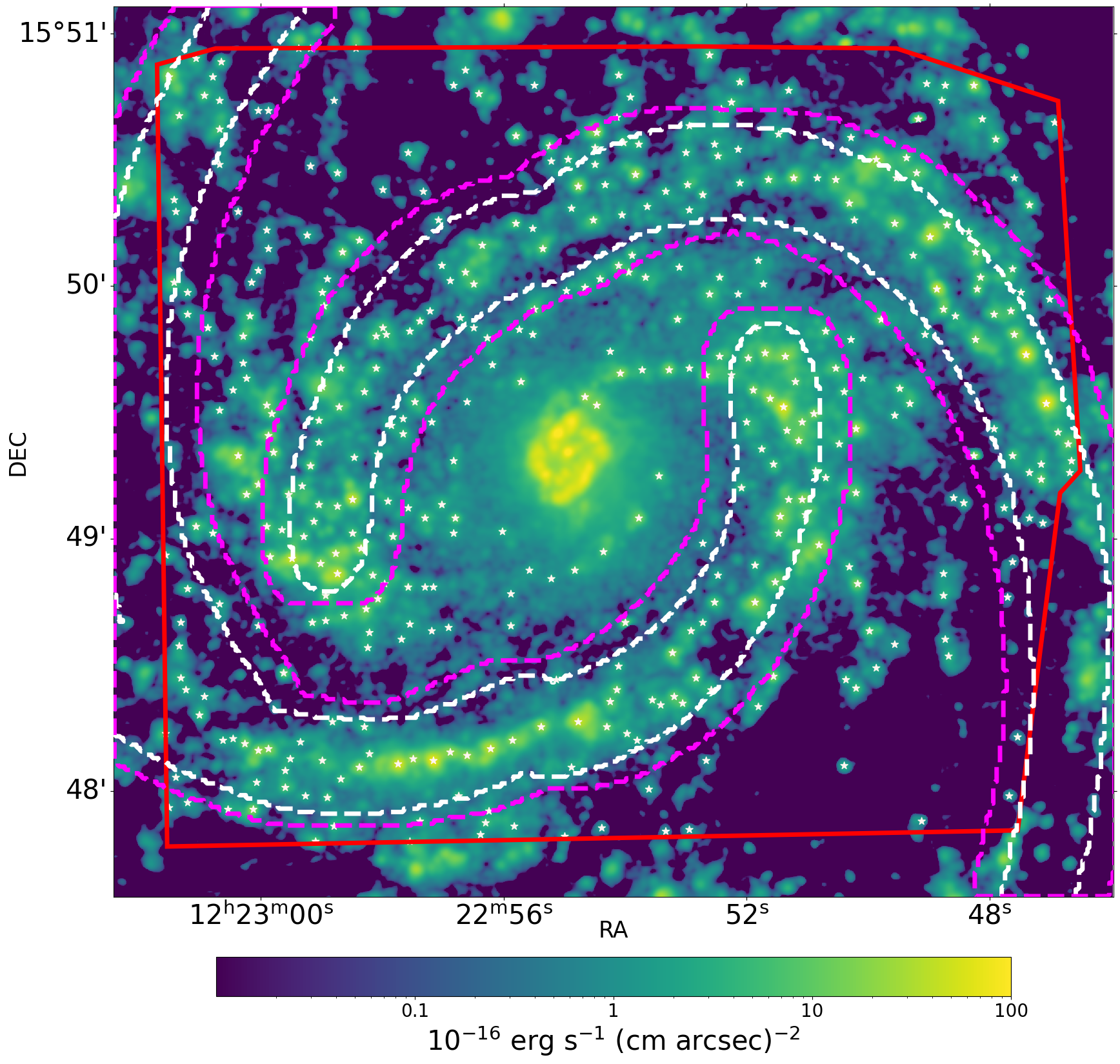}
\caption{Example of the environmental masks for NGC4321, overlaid on the \Halpha\ intensity map (Razza et al.\ in prep.). The width of the original spiral arm mask (white contours) from \citet{querejeta21} is adjusted by convolution with a Gaussian kernel (dashed pink contours) in order to visually encompass all the relevant \Halpha\ emission from the spiral structures. The red contour outlines the field-of-view of the ALMA CO(2-1) map, which limits the region in which our analysis can be performed. White stars mark the position of the identified emission peaks in the \Halpha\ map with the \textsc{Heisenberg} code (see Section~\ref{method} and \citealt{kim22}).}
\label{masks}
\end{figure}

The original masks defined by \citet{querejeta21} are optimised to specifically capture infrared and molecular gas emission. As a result, for some individual galaxies we enlarge the width of the spiral arm mask in order to also visually encompass all of the H$\alpha$ emission associated with spiral arm structures. To do this, we convolve the original mask with a Gaussian kernel, whose radius was varied to find an optimal value. The optimisation was done by visual inspection of the new mask, overlaid on the H$\alpha$ map. An example of the final mask is shown in Fig.~\ref{masks} for NGC~4321. 
We show in Fig.~\ref{ksrelation} the average molecular gas surface density as a function of the average SFR surface density in both environments for each galaxy. These observations follow the well-characterised Kennicutt–Schmidt relation \citep[e.g.\ ][]{schmidt59, kennicutt98}.

\subsection{Other data products}

\subsubsection{Global star formation rate maps} 
\label{SFR}
\begin{figure}[tbp]
\includegraphics[width=\columnwidth]{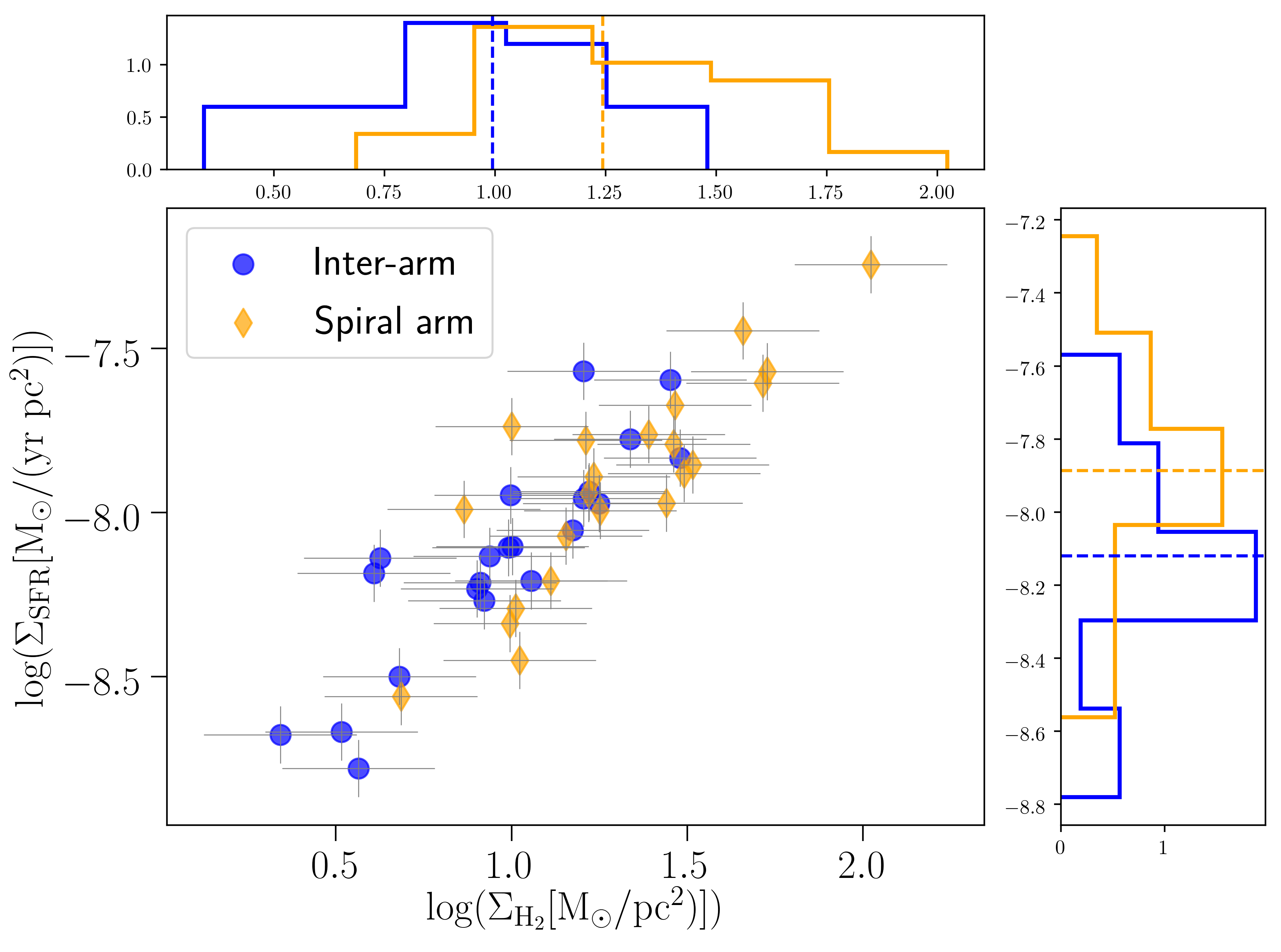}
\caption{Molecular gas surface density versus SFR surface density averaged over the area enclosed by arm and inter-arm regions. Dashed lines mark the position of the median of each sample.}
\label{ksrelation}
\end{figure}

H$\alpha$ emission from young, embedded star-forming regions can suffer from extinction. In order to recover the global extinction-corrected star formation rate of each galaxy, we use SFR maps built from a combination of multi-wavelength observations from the Galaxy Evolution Explorer (GALEX) in the far-UV and the Wide-field Infrared Survey Explorer (WISE) W4 band at $22~\mu$m maps \citep{leroy19}, convolved to 15\arcsec. These maps are used to estimate the global SFR density of Fig. \ref{ksrelation} and the global depletion timescale (defined in Sect.~\ref{SFE}).
\subsubsection{Metallicity maps}
\label{metallicity}

Gas metallicity enters into two important aspects of our analysis. Firstly, the CO-to-\Htwo\ conversion factor (see Sect.~\ref{alpha_CO}), necessary to estimate the total mass of molecular gas and therefore estimate the SFE, scales inversely with metallicity. Secondly, the timescales of the successive phases of the cloud-to-star evolutionary cycle derived using \textsc{Heisenberg} are fundamentally relative timescales, and the duration of one of these phases needs to be known to translate all the timescales to absolute ones \citep[the reference timescale in][]{kruijssen18}. \citet{haydon20} show that this reference timescale is weakly dependent on metallicity (also see Sect.~\ref{method}).

To recover the appropriate CO-to-\Htwo\ conversion factor and reference timescale in each environment, we use the PHANGS two-dimensional metallicity distribution \citep{williams22}. These metallicity maps are generated at a spatial resolution of 120\,pc for the 19 galaxies in the PHANGS sample observed with MUSE \citep{emsellem22}, using the S-calibration by \citet{pilyugin16}; of these, 13 belong to our sample. For the galaxies in our sub-sample that do not overlap with PHANGS-MUSE, we use the metallicity gradients measured by \citet{pilyugin14} when available (which is the case for NGC 1097, NGC 2997, NGC 5248, and NGC 6744). Finally, when no direct measurement is available, we adopt the metallicity from the global stellar mass-metallicity relation \citep{sanchez19} at the effective radius $R_{\rm eff}$ and assume a fixed radial metallicity gradient of $-0.1 {\rm dex}/R_{\rm eff}$ within the galaxy \citep{sanchez14}. We then calculate the CO-weighted average metallicity in each environment, defined by the masks described in Sect. \ref{envmask}.

\subsubsection{$\alpha_{\rm CO}$ conversion factor maps}
\label{alpha_CO}

To convert the observed CO flux to a gas mass, we adopt a metallicity-dependent $\alpha_{\rm CO}$ conversion factor, following \citet{sun20}:
\begin{equation}
\alpha_{\rm CO} = 4.35 \times Z^{-1.6}\ \rm M_{\odot}(K\ km\ s^{-1}\ pc^{2})^{-1},
\end{equation}
where $Z = (\rm O/H)$ is the oxygen abundance and $12+\log(\rm O/H)_{\odot} = 8.69$ \citep{asplund09} is the adopted solar value of reference. We estimate the gas mass surface density pixel by pixel as
\begin{equation}
    \Sigma_{\rm H_{2}} = \alpha_{\rm CO} \times R_{21}^{-1}\ I_{\rm CO}\ \cos(i),
\end{equation}
where $R_{21} = 0.65$ is the CO(2–1)-to-CO(1–0) line ratio \citep{leroy13, denbrok21}, $I_{\rm CO}$ is the intensity from the moment-0 ALMA-CO maps, and $i$ is the inclination of the galaxy \citep{lang20}. The $\cos(i)$ is needed to correct the derived surface densities to face-on projection, following \citet{sun22}.

Total gas masses are estimated pixel by pixel as
\begin{equation}
    M_{\rm H_{2}} = \alpha_{\rm CO} \times R_{21}^{-1}\ I_{\rm CO}\ A_{\rm pix}.
\end{equation}
In this case no correction for inclination is necessary, because both $I_{\rm CO}$ and $A_{\rm pix}$ are defined in the projected space. We note that we account for a conservative 50\% error on the total conversion factor, $\alpha_{\rm CO} \times R_{21}^{-1}$, given the uncertainties on these two parameters.

\section{Method} \label{method}

In this section, we describe our analysis method implemented in the \textsc{Heisenberg} code. More detail on theoretical aspects, code implementation and other observational applications can be found in \citet{kruijssen14, kruijssen18, kruijssen19, hygate19, chevance20, chevance22, haydon20, haydon20_ext, ward20_HI, ward22, zabel20, kim21, kim22, kim23, lu22}.

GMCs and star formation regions are tightly correlated on galactic scales, following the well-known 'star formation relation' \citep{kennicutt98}. However, observations at higher resolution ($\lesssim 1$\,kpc) provide a less tight correlation between SFR and gas surface density, with a large scatter of the 'resolved' relation around the 'global' one  \citep[e.g.][]{bigiel08, onodera10, feldmann11}. This scatter is a manifestation of the independent and rapid evolution of GMCs and star-forming regions, driven primarily by stellar feedback rather than by dynamical drift \citep[e.g.][]{kruijssen19,chevance20}. More precisely, the spatial decorrelation between GMCs and star-forming regions at small scales within galaxies arises as a consequence of resolving out different phases of the GMC-to-star life cycle \citep{kruijssen14}, with a GMC assembly phase, a star formation phase, and eventually a GMC disruption phase due to stellar feedback. This concept has been formalised as the 'Uncertainty Principle for Star Formation' \citep{kruijssen14} and has been provided as a publicly available framework in the form of the \textsc{Heisenberg} code\footnote{\href{https://github.com/mustang-project/heisenberg}{https://github.com/mustang-project/heisenberg}} \citep{kruijssen18}. We note that, while \citet{egusa04, egusa09} proposed that the observed spatial offset between CO and \Halpha peaks could be interpreted as a consequence of the movement of the spiral arm with respect to the gas and used directly to estimate an evolutionary timescale, this interpretation is not generalisable, because the offsets could be driven by a combination of cloud evolution, stellar feedback, and dynamical drift.

To translate this spatial decorrelation into the evolutionary timeline of GMCs and star formation, we first identified emission peaks in gas (here CO(2-1) emission) and SFR tracer (here H$\alpha$ emission) maps using the CLUMPFIND algorithm \citep{williams94}. We defined apertures of various sizes ranging between the spatial resolution of the coarser map, $l_{\rm ap,min}$, and a scale large enough to recover the correlation between the gas and stellar flux at the galactic scale (fixed for both spiral arms and inter-arms at approximately 3000\,pc) and place them around each peak to measure the relative change of gas-to-SFR flux ratio in a given aperture compared to the galactic average. This relative change as a function of the aperture size was then fitted by an analytical function that depends on three independent parameters: two timescales describing successive phases of the GMC life cycle and the characteristic separation distance.

The timescales of the successive phases of this GMC-to-star life cycle derived from this model are strictly speaking relative timescales (duration of one phase relative to another). A reference timescale is therefore needed to obtain absolute timescales from the measured variations of the gas-to-SFR flux ratios. As in previous studies, we calibrated the timeline by providing a reference timescale for the \Halpha-emitting phase of a simple stellar population ($t_{\rm star, ref}\sim4.3$\,Myr; \citealt{haydon20}). We accounted for the variation of $t_{\rm star, ref}$ with the gas-phase metallicity as:
\begin{equation}
t_{\rm star,ref} = (4.32\pm 0.16\,{\rm Myr})\times \left(\frac{Z}{{\rm Z}_{\odot }}\right)^{-0.086\pm 0.017}.
\end{equation}
In order to obtain the reference timescale for each environment, we applied the environmental masks to the metallicity maps, and we computed $t_{\rm star, ref}$ pixel by pixel. We then took the CO luminosity-weighted average as a best estimate of the reference timescale for each region. For our sample, each galaxy shows slight variations of metallicity between its arms and inter-arm regions. However, when all the galaxies are grouped together, the range of metallicity values (and therefore of reference timescales) is similar for arms and inter-arm regions. The metallicity is distributed within a range of $log(O/H)+12 = [8.41;8.65]$ both for arms and inter-arms, and the reference timescale $t_{\rm star,ref}$ also spans a very limited range of $[4.30;4.59]$\,Myr. There is therefore no systematic change in timescales between arms and inter-arm regions due to metallicity variations (see Table~\ref{table1}).

With this reference timescale, we can translate the relative duration of each phase into an absolute one. We obtain: the cloud lifetime (the timescale over which CO is visible, denoted as $t_{\rm CO}$), the feedback timescale (during which both CO and H$\alpha$ are co-spatial, $t_{\rm fb}$),\footnote{Multi-scale analysis of this co-spatial phase shows that it is indeed the disruptive effect of stellar feedback that causes the separation of CO and H$\alpha$, rather than a form of spatial drift \citep{kruijssen24}.} and $\lambda$, the characteristic separation length between independent regions. From these three main parameters, we also derive other physical quantities of interest for this study, such as the depletion time for compact sources, the SFE, and the fraction of diffuse molecular gas (see Sect.~\ref{results}).

The presence of large-scale, diffuse emission affects the measured cloud lifetime, because it includes emission that originates from unbound structures that do not participate in the GMC-star evolutionary cycle \citep{kruijssen18,hygate19}. Therefore, diffuse emission on scales much larger than the typical distance between independent regions must be filtered from the observed tracer maps to ensure an accurate estimate of the timescales characterising the cloud life cycle. The origin of this diffuse emission is very heterogeneous and depends on the tracers used. As a result, many different criteria exist in the literature to filter it out. Here, we took an agnostic, physics-based approach by filtering the large-scale emission from both CO(2-1) and \Halpha\ maps using the method presented in \citet{hygate19}. We used a Gaussian high-pass filter and filtered out the (spatial) frequencies corresponding to a spatial scale larger than a multiple ($n_{\rm \lambda}\gtrsim 10$) of the mean separation length ($\lambda$) in Fourier space (see Table 2 in \citealt{kim22}). This process was applied iteratively, so we did not make assumptions on the spatial scale at which the filtering is done. The filtering of the diffuse emission and the fit of the gas-to-stellar flux ratio as a function of spatial scale were repeated, until the derived separation scale $\lambda$ did not vary for four successive iterations. Because we filtered the diffuse large-scale emission of gas, the derived physical quantities measured, such as the molecular gas surface density, the depletion timescale, and the integrated SFE derived in Sect.~\ref{results}, refer to the compact emission only arising from GMCs.

We performed our statistical analysis to each of the 22 galaxies. For each galaxy, the main input parameters are listed in Table 1 of \citealt{kim22}. Default values from \citet{kruijssen18} were used for other fitting and error propagation parameters. Since we used the metallicity maps from \citet{williams22}, we also updated the results from \citep{kim22}, using the appropriate metallicity-dependent reference timescale for consistency. This results in variations of $t_{\rm ref}$ (and therefore of $t_{\rm CO}$ and $t_{\rm fb}$) of $~5\%$ on average, which stays comfortably within the 1-sigma uncertainties reported by \citep{kim22}. In the following, figures showing results from \citep{kim22} have been updated accordingly.

We selected galactic substructures by masking each map with the environmental masks described in Sect.~\ref{envmask}, in order to separate arm and inter-arm regions. On these substructures, we ran the \textsc{Heisenberg} code each time to extract the parameters describing the evolutionary timeline of their clouds. We made sure that the galactic substructures from the masks had a minimum width of 1\,kpc so that the regions were at least $>2\lambda$ in size. This ensured that the realisation of the independent star-forming regions would approximate a random normal distribution \citep{kruijssen18}. The choice of such broad masks, each encompassing $\gtrsim 100$ regions, considerably limits the number of individual regions being exactly split between two different environments to a very small fraction. In the few cases where this might have occurred, this did not bias our results. Slightly increasing or decreasing the size of the masks leads to similar measurements of the cloud lifetime, feedback timescale, and $\lambda$.

\section{Results}
\label{results}

In this section we present the results obtained by performing our statistical analysis \citep{kruijssen14} on 22 spiral galaxies. In Fig. \ref{ksrelation} we show the Kennicut-Schmidt relation, where SFR and the molecular gas mass have been measured by summing up all the emission in the area effectively covered by the masks before performing any filtering for diffuse emission. We measured the emission of CO and \Halpha\ peaks and fitted the underlying bias following the method detailed in Sect.~\ref{method}. In order to characterise the environmental dependence of the cloud life cycle, we applied the method in spiral structures and inter-arm regions for each galaxy. The results are presented in Table~\ref{table1}. 

\begin{figure*}[tbp]
\includegraphics[width=\textwidth]{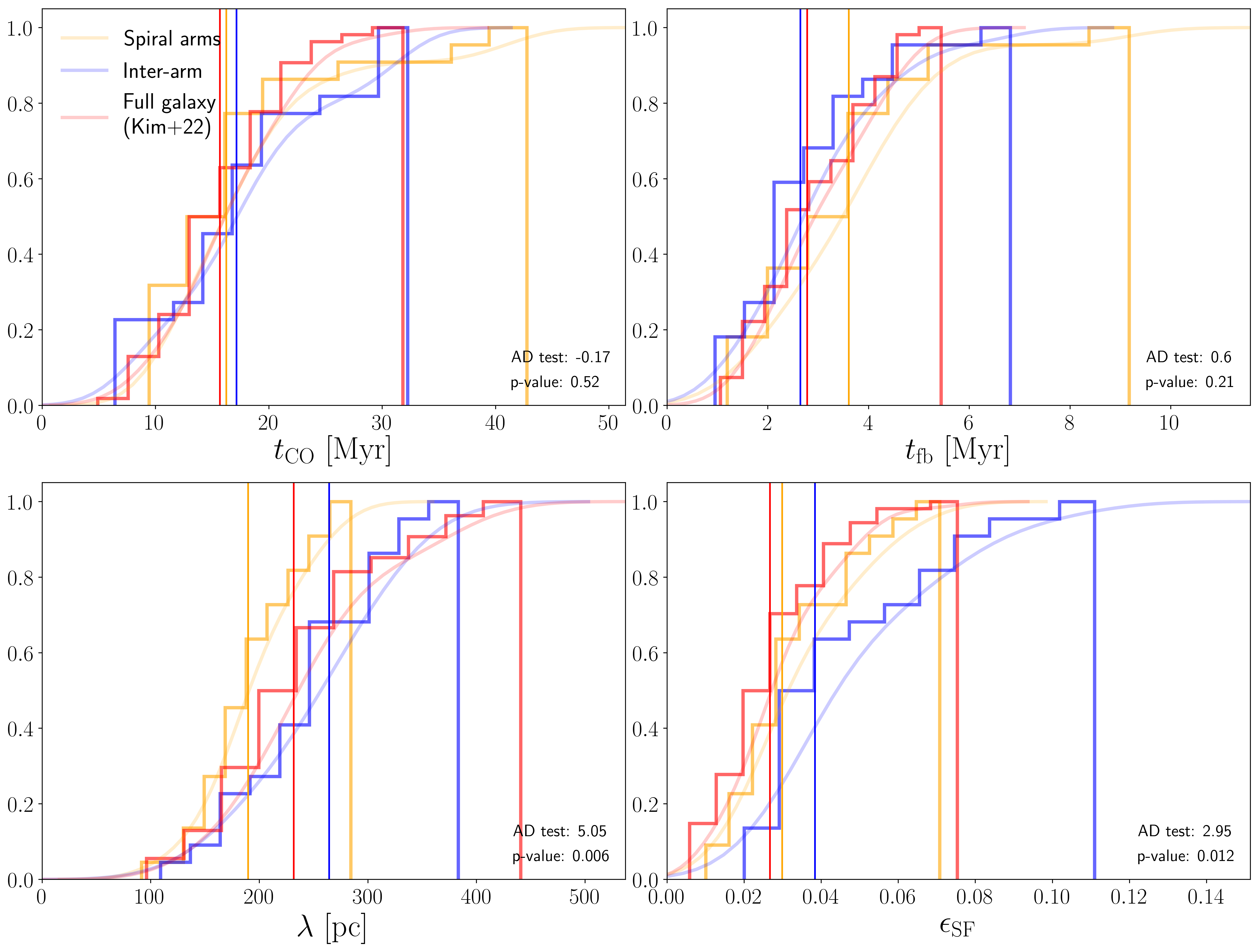}
\caption{Cumulative distribution function of the cloud lifetime (top left), feedback timescale (top right), region separation length (bottom left), and SFE (bottom right). Each panel shows the c.d.f.\ for the parameters calculated in spiral arms (orange) and inter-arm regions (blue), as well as for the full galaxies (red, \citealt{kim22}) as a thick solid line, and the smoothed c.d.f as a shaded line. In cases where only an upper limit could be determined, the value indicated in Table~\ref{table1} is used.} Medians are indicated with solid vertical lines. Anderson-Darling test statistics and $p$-values are indicated in the bottom right of each panel.
\label{cdf}
\end{figure*}

Figure~\ref{cdf} shows the cumulative distribution functions (c.d.f.) of the three independent parameters that characterise the cloud life cycle (the cloud lifetime, $t_{\rm CO}$; the feedback timescale, $t_{\rm fb}$; and the region separation length, $\lambda$), and of the SFE ($\varepsilon_{\rm SF}$).
The distributions were estimated using a Kernel Density Estimation algorithm, with a Gaussian kernel, provided by the Python module \textsc{statmodels}. For each of these parameters, we assessed whether the two cumulative distributions differ when split by environment, by conducting a k-sample Anderson-Darling (AD) test. This modified AD test \citep{Anderson54,scholz87} tests the null hypothesis that different samples are drawn from the same population.

\begin{figure*}[tbp]
\includegraphics[width=\textwidth]{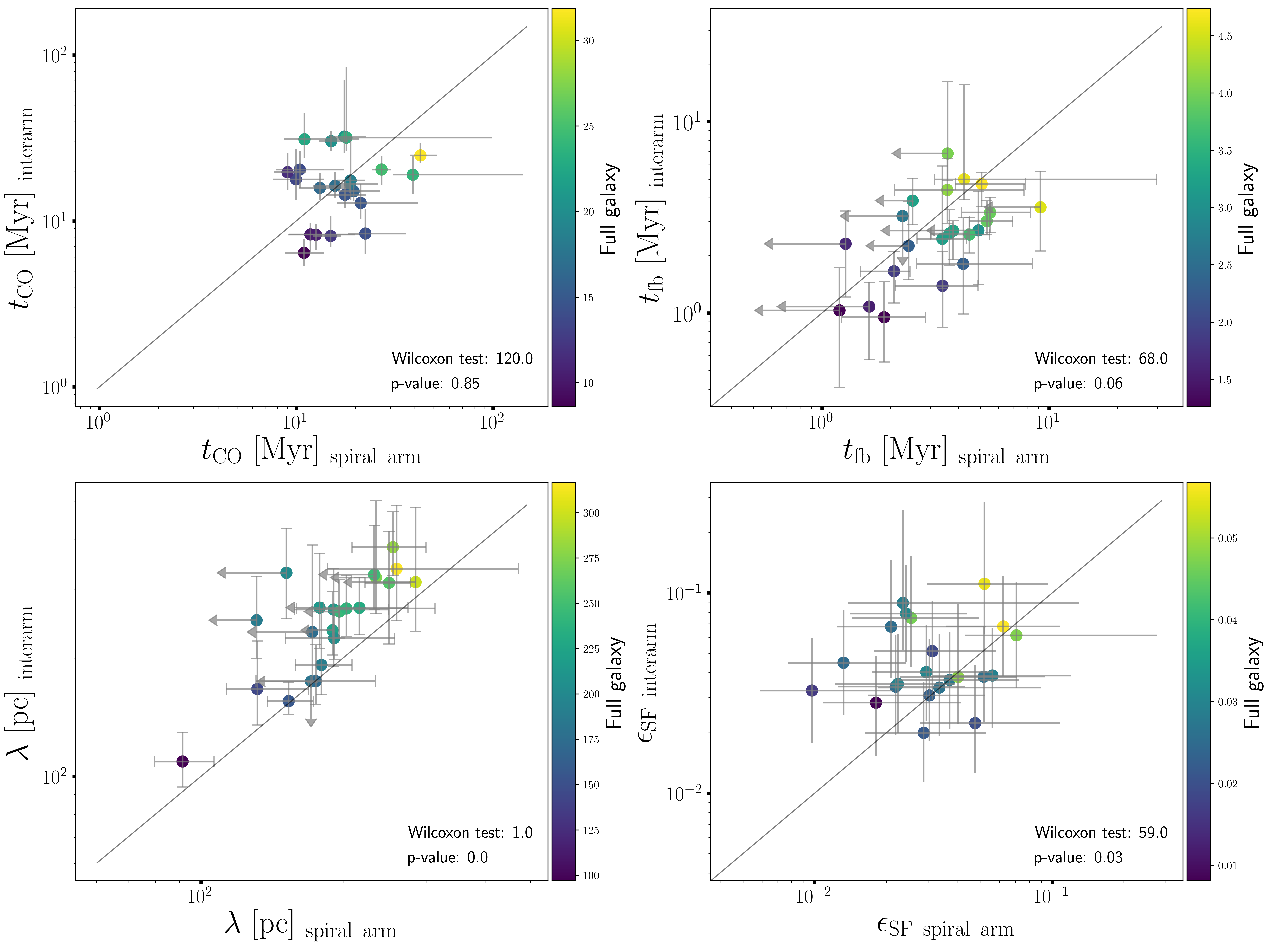}
\caption{Cloud lifetime (top left), feedback timescale (top right), region separation length (bottom left), and integrated SFE (bottom right) measured through our statistical analysis. For each parameter, we report the values calculated in the spiral arms on the x-axis and in the inter-arm regions on the y-axis for each galaxy. The grey line shows the one-to-one relation. Each data point is colour-coded according to the averaged value obtained by \citet{kim22} for the full galaxy. For the feedback timescale and the region separation length, upper limits in the spiral arms are indicated with arrows.}
\label{scatter}
\end{figure*}

In addition, we also show for each parameter in Fig.~\ref{scatter}, the comparison between the value in the arms and in the inter-arm regions for each galaxy.
Overall, we do not see any systematic shift from the one-to-one relation as a function of the galactic averaged values obtained in \citet{kim22}. We used the Wilcoxon signed-rank test \citep{Wilcoxon45} for paired samples to check any differences in the estimated parameter median. This is a non-parametric alternative to the Student $t$-test that does not assume normality for the distribution of the errors. It is a natural choice to use paired sample statistics in this work because of the way the two samples (spiral arms and inter-arm regions) are defined. We report in Table~\ref{table2} the statistics and $p$-values associated with both tests. We caution that, when only an upper limit could be determined for $\lambda$ and $t_{\rm fb}$, we used this upper limit as our best-guess value to perform these tests and compute the statistics and associated p-values, in order to maximise the sample size.
\begin{table}[!htbp]
\centering
\begin{tabular}{c c c c c} 
\hline
\hline
 & $t_{\rm CO}$ & $t_{\rm fb}$ & $\lambda$ & $\epsilon_{\rm SF}$ \\ [0.5ex] 
 \hline
 \rule[-4mm]{0mm}{1cm}
 AD test & -0.17 & 0.60 & 5.05 & 2.95 \\ 
 $p$-value & 0.52 & 0.21 & 0.006 & 0.012 \\ [1ex] 
 \hline
 \rule[-4mm]{0mm}{1cm}
 Wilcoxon test & 120.0 & 68.0 & 1.00 & 59.0 \\ 
 $p$-value & 0.85 & 0.06 & 0.00 & 0.03 \\ [1ex] 
 \hline
 \end{tabular}
\caption{Results of the AD and Wilcoxon tests with the associated $p$-values between spiral arms and inter-arm regions for the cloud lifetime ($t_{\rm CO}$), the feedback timescale ($t_{\rm fb}$), the region separation length ($\lambda$), and the integrated SFE ($\epsilon_{\rm SF}$).}
\label{table2}
\end{table}
In the following sections, we discuss in more detail the results of these statistical tests for each of the parameters shown in Figs.~\ref{cdf} and \ref{scatter}.

\subsection{Cloud lifetime ($t_{\rm CO}$)}
Across all the galaxies, the measured cloud lifetimes in both environments fall within the range 5 to 40\,Myr, which is in very good agreement with previous applications of the statistical method used here to full galaxies \citep{kruijssen19, chevance20, kim22}, as well as the cloud lifetimes estimated with other techniques \citep{kawamura09, Corbelli17}. The cumulative distributions of the cloud lifetime in spiral and inter-arm regions do not show statistically significant differences according to the AD test, with a $p$-value of 0.52 (Fig.~\ref{cdf}). In addition, the scatter around the one-to-one line (Fig. \ref{scatter}) is very broad (0.25dex), with a root mean square deviation (RMSD) and maximum absolute error (MAE) equal to 2.2~Myrs and 8.1~Myrs respectively, but there is neither a systematic shift towards a preferred environment, nor any trend with the globally averaged value measured by \citet{kim22}.  Indeed, the Wilcoxon test does not capture any difference in the medians of the two distributions ($p$-value $>0.05$), meaning that these two populations are statistically indistinguishable.

\subsection{Feedback timescale ($t_{\rm fb}$)}
The feedback timescale for both spiral arm and inter-arm regions span a range of 1-10\,Myr, which is in agreement with previous studies for full galaxies \citep{kruijssen19, chevance20, kim22}. According to the AD test, the difference between the feedback timescale distributions in spiral arm and inter-arm regions is not statistically significant, with a $p$-value of $0.21$. However, we note that the AD test checks for overall differences in the cumulative distributions, and, as seen in Fig.~\ref{cdf}, the feedback timescale shows some variation in the range of values spanned by both environments, with inter-arm regions having on average lower values than spiral arm regions. First, the scatter between galaxies is tighter than for the cloud lifetime (0.20 dex). The RMSD and MAE are equal to 0.38~Myrs and 1.41~Myrs, respectively. Secondly, there is a small but significant difference between the two medians,  which is well captured by the Wilcoxon signed-rank test with a $p$-value 0.06 (see Fig.~\ref{scatter}). However, we note that this difference might decrease if exact measurements were obtained instead of upper limits for some spiral arms. 

\subsection{Region separation length ($\lambda$)}
The average separation length between independent star-forming regions ($\lambda$, bottom left panel in Fig. \ref{cdf}) is $259\pm79$\,pc, when considering entire galaxies \citep{kim22}. The separation distance is defined as the scale at which the gas-to-stellar flux in the apertures focusing on the gas or stellar peaks significantly deviates from the galactic average. As specified in \citet{chevance20}, $\lambda$ describes the length scale in the immediate vicinity of a region over which a sufficiently large number of neighbouring regions is found to wash out the spatial decorrelation between the gas and stellar flux seen on small scales. This scale length is therefore a local measurement and is not affected by the area of the covered field-of-view. This can be further seen by comparing $\lambda$ measured by \citet{kim21} on the whole disc with respect to what we obtained in this study: $\lambda$ in inter-arms and spiral arms measured in the current work are respectively larger and smaller than the full galaxy disc estimate, despite both analysed fields-of-view being smaller than the full galaxy (by construction).

When splitting between environments, the separation length spans a range of 100-300\,pc in spiral arms, with a median of ${\sim} 175$\,pc, and a much wider range of values (up to 400\,pc) in inter-arm regions, with a higher median (${\sim}250$\,pc). The scatter around the one-to-one line is 0.08 dex, which is much smaller than for the timescales described above. The statistically significant difference between the two distributions is confirmed by the AD and the Wilcoxon tests ($p$-values $< 0.05$). Specifically, for the AD test, the null hypothesis (i.e. the two samples coming from the same distribution) can be rejected at the 0.5\% confidence level \citep{scholz87}. This result would be strengthened even further by obtaining exact measurements instead of upper limits for $\lambda$ in some of the spiral arms.
This result is expected when doing a simple visual inspection of the tracer maps and of the distribution of the emission peaks, as in Fig.~\ref{masks}, where the emission peaks are more sparsely located in the inter-arm region than in the spiral arm. This is also in agreement with the lower gas surface density and lower SFR surface density measured in the inter-arm regions of spiral galaxies compared to the spiral arm regions \citep[e.g.][]{Vogel1988, GarciaBurillo1993, NakanishiSofue2003, Hitschfeld2009, sun20, querejeta21,querejeta24}.

\subsection{Star formation efficiency ($\epsilon_{\rm SF}$)}
\label{SFE}
Following \cite{kruijssen18}, we calculated the integrated SFE per cloud as:
\begin{equation}
    \epsilon_{\rm SF} = \frac{t_{\rm CO}}{t_{\rm dep}^{\rm comp}},
    \label{eq:SFE}
\end{equation}
where \(t_{\rm dep}^{\rm comp} = \Sigma_{\rm H_{2}}^{\rm comp} / \Sigma_{\rm SFR}\)
is the depletion time-scale of compact molecular gas structures (clouds), assuming that all star formation takes place in such structures (after removing the diffuse CO emission, see Sect.~\ref{method})\footnote{We note that the (commonly used) SFE per free-fall time is different than the integrated SFE per star formation event measured here, following Eq.~\ref{eq:SFE} by two aspects. First, we considered only the compact gas component that actually participates in the star formation process when estimating the depletion time. Second, we integrated the SFE per cloud lifetime ($t_{\rm CO}$), which is typically a few times the free-fall time, with considerable scatter depending on the environment (see e.g. \citealt{kruijssen19, chevance20, chevance23})}. We used the global SFR maps (as defined in Sect. \ref{SFR}) without filtering the diffuse emission. The diffuse \Halpha\ mostly comes from photons leaking out of \HII\ regions \citep[e.g.][]{Mathis1986, Sembach2000, Wood2010, belfiore22}; therefore, it had to be accounted for in the total star formation budget. The SFE ranges between 1-7\% in the case of spiral arm regions, with a median of $\sim 2.9$, while it tends to be on average slightly higher in the inter-arm regions (2-11\%), with a median of $\sim 3.8$. These findings are in agreement with \citet{kim22}, who find the $\epsilon_{\rm SF}$ between 1-8\%, with a mean of $2.9$ in galaxies of the PHANGS sample. The scatter of the SFE between the two environments around the one-to-one line is 0.27 dex.

According to the AD test, the two distributions are statistically different, as shown by the $p$-value of $\sim 0.012$, with inter-arm regions having on average higher star formation efficiencies than spiral arms regions. We can reject the null hypothesis at the 1\% confidence level \citep{scholz87}. This can be directly understood from fact that the cloud lifetime does not systematically differ between the two environments, while the depletion timescale appears to be systematically longer in spiral arms (as already visible in Fig. \ref{ksrelation}. If they exist, systematic differences in the conversion factors $R_{21}$ or $\alpha_{\rm CO}$ between arms and inter-arm could affect this result, but there is currently no consensus on this point in the literature (see e.g.\ \citealt{Koda2012, denBrok22}). We discuss this result in comparison with other measurements from the literature in Sect.~\ref{discussion}.

\section{Discussion}
\label{discussion}

We applied the 'Uncertainty principle for star formation' to 22 spiral galaxies in the PHANGS sample. We used H$\alpha$ and CO emission line maps to trace respectively the star-forming and molecular gas phases of the matter cycle in galaxies. We measured the relevant timescales characterising this cycle in both the spiral arms and inter-arm regions separately using environmental masks in order to assess the role of spiral arms in the star formation process. 

Our main finding is that molecular clouds live for the same amount of time, independently from their location (within spiral arm or in inter-arm regions). 
However, there are statistically significant differences in the spatial distribution of the independent regions undergoing this evolutionary cycle between gas clouds and star-forming regions: the separation length between regions is shorter by ${\sim} 100$\,pc in spiral arms than in inter-arm regions. Therefore, molecular clouds and star-forming regions are on average closer together in spiral arms. This result is expected, as spiral arms typically show a higher concentration of emission peaks. The timescale over which stellar feedback disrupts the parent molecular clouds shows similar distributions in both environments, as revealed by the AD statistics. However, we note that the median values of the two distributions differ significantly, as shown by the Wilcoxon rank-signed test. While we do not expect a significant impact of extinction at the moderate molecular gas densities present in our sample, the enhanced gas density in spiral arms compared to inter-arm regions (Fig.~\ref{ksrelation}) could lead to overestimated timescales in the arms \citep{haydon20_ext} in cases where $\Sigma_{ \rm mol}>20$M$_{\odot}$\,pc$^{-2}$. At low levels of extinction, corresponding to mean galactic surface densities $\Sigma_{\rm mol}<20M_{\odot}pc^{-2}$, \citet{haydon20_ext} find a slight increase in the timescales of the extinct map compared to the non-extinct one; however, the results remain consistent within the 1-sigma error range. Correcting for extinction could therefore slightly increase the observed difference between the median feedback timescales in arms and inter-arms, while in some extreme cases, it could erase the slight difference. Additionally, we observe a statistically significant difference in integrated star formation efficiencies between both environments, with spiral arms having $[25,50,75]^{\rm th}$ percentiles of $[2.2,2.9,4.5]\%$ in converting gas, while inter-arms have percentiles equal to $[3.3,3.8,6.6]\%$.

The picture of GMC evolution arising from our study is that clouds in spiral arms live for a similar amount of time and form stars with a slightly lower efficiency as clouds in the inter-arms.\footnote{While we neither define GMCs as bound nor even static objects, it might be useful to broadly characterise the units that undergo the evolutionary cycle identified in this work for reference. 
We extracted the properties of the emission peaks identified with CLUMPFIND and modelled as Gaussian profiles, as described in \citet{kruijssen18}. Across all the galaxies in our sample, we identify a total of 11,158 CO peaks, with masses $\log(M[\rm M_{\odot}]) = [5.17,5.57,6.06]$ (with a lower limit consistent with the sensitivity limit at $90\%$ sample completeness of $10^5$\,M$\odot$; \citealt{Leroy2021a}) and radii $R[\rm pc] = [57,79,110]$, corresponding to the [25,50,75]$^{\rm th}$ percentiles of the distribution.
Slightly larger values are found by \citet{rosolowsky21} using the cloud characterisation algorithm CPROPS, with $\log(M[\rm M_{\odot}]) = [5.92,6.28,6.67]$ and $ R[\rm pc] = [86,118,152]$ (corresponding to the [25,50,75]$^{\rm th}$ percentiles,  similar in both spiral arms and inter-arm regions). This is due to the fact that the algorithm in some cases identifies GMC complexes of several 100 pc in size (larger than $\lambda$). See \citet{chevance20} for a more detailed comparison.}
This is a strong indication against the suggestion that spiral arms might trigger star formation in GMCs. However, GMCs are more closely packed in spiral arms. The concentration of clouds in spiral arms, in combination with a higher gas surface density altogether, results in a lower surface density contrast, $\varepsilon_{\rm CO}$, calculated as the ratio between the average gas mass surface density at the peak locations ($\Sigma_{\rm peak,CO}$) and the global gas mass surface density over the considered field-of-view for that specific environment ($\Sigma_{\rm global,CO}$):
\begin{equation}
\varepsilon_{\rm CO} = \Sigma_{\rm peak,CO}/\Sigma_{\rm global,CO}.
\end{equation}
As shown in \citet{kim22}, a sparse medium (high surface density contrast) facilitates the dispersal of the GMCs, while a denser medium (low surface density contrast) tends to result in a longer dispersal timescale. This is in agreement with our finding that the feedback timescale is on average slightly longer in the spiral arms relative to the inter-arm.

\begin{figure}[tbp]
\includegraphics[width=\columnwidth]{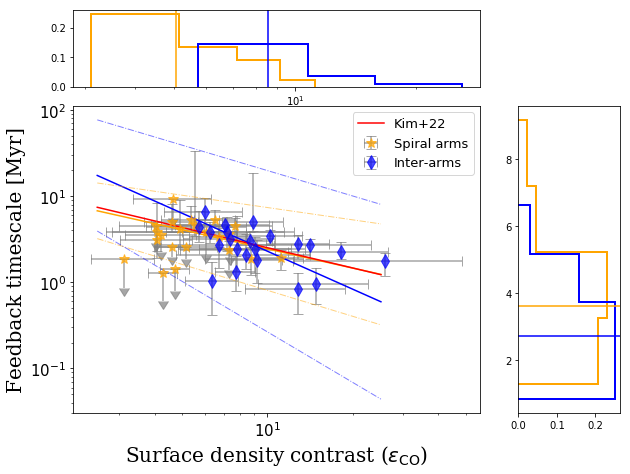}
\caption{Scatter plot of the feedback timescale as a function of the surface density contrast between the CO emission peaks and the galactic average ($\varepsilon_{\rm CO} $; see text). Spiral arms are represented in orange, inter-arm regions in blue. The solid lines represent the linear models fitted to the data in logarithmic space for each of the two environments. The red line marks the correlation obtained by \citet{kim22} for full galaxies, on a wider sample of 54 galaxies (including the 22 analysed in this work). The dashed lines represent the $1\sigma$ confidence intervals of the linear model for spiral arms (orange) and inter-arm regions (blue).}
\label{contrast}
\end{figure}

To verify this hypothesis, we show in Fig.~\ref{contrast} the correlation between the feedback timescale and the gas surface density contrast.
We indeed confirm that the lower gas surface density of inter-arm regions (as already shown in Fig.~\ref{ksrelation}) is accompanied by a higher $\varepsilon_{\rm CO}$, denoting a sparse environment. Most of the CO emission appears to be concentrated within the GMCs, with little molecular gas around them. By comparison, in spiral arms the density contrast is lower, implying that gas peaks are surrounded by a denser medium. Within both environments (arms and inter-arms independently), we observe a decrease in the feedback timescale with increasing surface density contrast, following the trend identified by \citet{kim22}. This seems to support the idea that spiral arms collect molecular gas but do not drastically change the process of star formation within the GMCs.
Clouds live for a similar amount of time, and star formation occurs within a dynamical time \citep{chevance20}, without being specifically triggered by spiral arm passages. Indirectly, the higher gas surface density accumulated in spiral arms is accompanied by a globally denser medium, where it takes more time to disperse clouds than in the sparse medium of inter-arm regions. This leads to the observed increase in the feedback timescale in spiral arms.

Figures~\ref{cdf} and~\ref{scatter} show that the integrated SFE per star formation event is $~40-50\%$ larger in the inter-arm regions than in the arm itself. This result seems to be in tension with the fact that the instantaneous SFE is found to be similar in spiral arms and inter-arm regions for a sample of galaxies almost identical to the one studied here \citep{querejeta24} and even more with the increase in the instantaneous SFE in the arms suggested in several earlier studies \citep[e.g.][]{Vogel1988, cepa90, lord90, knapen96}. 
This discrepancy could be (partially) explained by the fact that the estimated $\epsilon_{\rm SF}$ in the inter-arms might be slightly overestimated by including ionising photons from star-forming regions in the arms leaking in the inter-arm environment (see Sect.~\ref{SFR}). Indeed, the mean free path of ionising photons is comparable or larger to the width of our environmental masks ( $>1$\,kpc; \citealt{belfiore22}). By contrast, photons leaking from star-forming regions in the inter-arms towards the spiral arm environment would only represent a marginal contribution to the measured SFR surface density in comparison to the local contribution. As a result, $\Sigma_{\rm SFR}$ (resp.\ $t_{\rm dep}^{\rm comp}$) could be slightly increased (respectively\ decreased) due to the contribution of large scale diffuse ionised gas \citep[see Appendix D]{querejeta24}. 
Other factors can potentially affect our measurement of the SFE. In particular, a systematic variation of the $\alpha_{\rm CO}$ or the $R_{21}$ conversion factors between spiral arms and inter-arms could modify our conclusion on $\epsilon_{\rm SF}$. However, this remains a debated topic, and differences in flux calibration of the observations can drive very different outcomes regarding how $R_{21}$ varies between arms and inter-arms, even on the same target \citep[e.g.\ in M51][]{Koda2012,denBrok22}. In the absence of a definitive answer, we chose the luminosity weighted averaged $R_{21}$ provided in \citep{denbrok21} across our full sample. The 50\% uncertainties in the light-to-mass conversion factor ($\alpha_{\rm CO} \times R_{21}^{-1}$; see Section~\ref{alpha_CO}), encompasses these systematics unknown in the mass estimate.

Despite these points of caution, our results are consistent with observations showing that clouds in spiral arms have on average lower virial parameters than clouds in the inter-arm regions \citep[e.g.\ ][]{hirota18, rosolowsky21}. Despite the slightly lower SFE in spiral arms than in inter-arm regions, the higher GMC masses \citep[e.g.\ ][]{hirota18, rosolowsky21} might explain the more massive, emerging young stellar clusters found in M83 by Knutas et al.\ (in prep.) \citep{kruijssen14c}. This is in contrast with theoretical prediction of the SFE per free-fall time measured at the cloud scale, for which it is expected to be anti-correlated  with the virial parameter \citep{krumholz05,padoan12}. However, as demonstrated in Fig.~\ref{scatter}, there is considerable spread in the SFE between different galaxies, and the generalisation of this result would need to be confirmed.

In conclusion, these observations do not provide evidence for a strong role of spiral arms in the star formation process in galaxies. Spiral arms likely have a major role in accumulating the gas \citep{querejeta21}, but the clouds forming inside them live on average for a similar amount of time as the ones in the inter-arm regions. The exact GMC lifetime in the arms can be longer or shorter than those in the inter-arm depending on the galaxy. They are dispersed by stellar feedback on a similar timescale (on average, a slightly longer one) and seem to form stars at a slightly lower efficiency in spiral arms. In the outskirts of grand design spiral galaxies, molecular clouds could be more strongly affected by spiral arm crossing as the gravitational stability and galactic shear support increase \citep{jeffreson18}. Even though the correct estimate of the virial parameter depends strongly on the assumptions made for the determination of the $\alpha_{\rm CO}$ conversion factor, on average the population of clouds in the inter-arm shows a slightly higher $\alpha_{\rm vir}$ with respect to those in spiral arms, regardless of the recipe used to convert CO light to molecular gas mass (\citealt{rosolowsky21}, Hughes et al.\ in prep.). This counter-intuitive result of less bounded clouds forming stars with higher efficiencies is also observed in \cite{Leroy2025}, who find a mild positive correlation between the SFE per free-fall time and the virial parameter measured on the cloud scale. This suggests that the virial parameter measured on $\sim100$\,pc scales, considering self-gravity only, does not fully account for the amount of self-gravitating, likely star-forming gas. A broader, more diffuse gas component bound partially by stellar gravity cannot be ignored to measure the true dynamical state of a cloud \citep[e.g.][]{meidt18, Liu2021}.

\begin{acknowledgements}
AR, MC and LR gratefully acknowledge funding from the DFG through an Emmy Noether Research Group (grant number CH2137/1-1).
JMDK gratefully acknowledges funding from the European Research Council (ERC) under the European Union's Horizon 2020 research and innovation programme via the ERC Starting Grant MUSTANG (grant agreement number 714907). 
COOL Research DAO \citep{cool_whitepaper} is a Decentralised Autonomous Organisation supporting research in astrophysics aimed at uncovering our cosmic origins. 
MQ acknowledges support from the Spanish grant PID2022-138560NB-I00, funded by MCIN/AEI/10.13039/501100011033/FEDER, EU.
J.K. is supported by a Kavli Fellowship at the Kavli Institute for Particle Astrophysics and Cosmology (KIPAC). 
J.S. acknowledges support by the National Aeronautics and Space Administration (NASA) through the NASA Hubble Fellowship grant HST-HF2-51544 awarded by the Space Telescope Science Institute (STScI), which is operated by the Association of Universities for Research in Astronomy, Inc., under contract NAS~5-26555.
JPe acknowledges support by the French Agence Nationale de la Recherche through the DAOISM grant ANR-21-CE31-0010 and by the Thematic Action “Physique et Chimie du Milieu Interstellaire” (PCMI) of INSU Programme National “Astro”, with contributions from CNRS Physique \& CNRS Chimie, CEA, and CNES.
SCOG acknowledges financial support from the European Research Council via the ERC Synergy Grant ``ECOGAL'' (project ID 855130) and from the German Excellence Strategy via the Heidelberg Cluster of Excellence (EXC 2181 - 390900948) ``STRUCTURES''.

This paper makes use of the following ALMA data: \linebreak
ADS/JAO.ALMA\#2012.1.00650.S, \linebreak
ADS/JAO.ALMA\#2013.1.01161.S, \linebreak 
ADS/JAO.ALMA\#2015.1.00925.S, \linebreak 
ADS/JAO.ALMA\#2015.1.00956.S, \linebreak 
ADS/JAO.ALMA\#2017.1.00886.L, \linebreak

ALMA is a partnership of ESO (representing its member states), NSF (USA) and NINS (Japan), together with NRC (Canada), MOST and ASIAA (Taiwan), and KASI (Republic of Korea), in cooperation with the Republic of Chile. The Joint ALMA Observatory is operated by ESO, AUI/NRAO and NAOJ. This paper includes data gathered with the 2.5 meter du Pont located at Las Campanas Observatory, Chile, and data based on observations carried out at the MPG 2.2m telescope on La Silla, Chile.
\end{acknowledgements}

\bibliographystyle{aa}
\bibliography{bibliography}

\begin{thebibliography}{115}
\expandafter\ifx\csname natexlab\endcsname\relax\def\natexlab#1{#1}\fi

\bibitem[{Anderson \& Darling(1954)}]{Anderson54}
Anderson, T.~W. \& Darling, D.~A. 1954, Journal of the American Statistical Association, 49, 765

\bibitem[{{Asplund} {et~al.}(2009){Asplund}, {Grevesse}, {Sauval}, \& {Scott}}]{asplund09}
{Asplund}, M., {Grevesse}, N., {Sauval}, A.~J., \& {Scott}, P. 2009, \araa, 47, 481

\bibitem[{{Belfiore} {et~al.}(2022){Belfiore}, {Santoro}, {Groves}, {Schinnerer}, {Kreckel}, {Glover}, {Klessen}, {Emsellem}, {Blanc}, {Congiu}, {Barnes}, {Boquien}, {Chevance}, {Dale}, {Diederik Kruijssen}, {Leroy}, {Pan}, {Pessa}, {Schruba}, \& {Williams}}]{belfiore22}
{Belfiore}, F., {Santoro}, F., {Groves}, B., {et~al.} 2022, \aap, 659, A26

\bibitem[{{Bigiel} {et~al.}(2008){Bigiel}, {Leroy}, {Walter}, {Brinks}, {de Blok}, {Madore}, \& {Thornley}}]{bigiel08}
{Bigiel}, F., {Leroy}, A., {Walter}, F., {et~al.} 2008, \aj, 136, 2846

\bibitem[{{Blitz}(1990)}]{Blitz1990}
{Blitz}, L. 1990, in Astronomical Society of the Pacific Conference Series, Vol.~12, The Evolution of the Interstellar Medium, ed. L.~{Blitz}, 273--289

\bibitem[{{Brinchmann} {et~al.}(2004){Brinchmann}, {Charlot}, {White}, {Tremonti}, {Kauffmann}, {Heckman}, \& {Brinkmann}}]{Brinchmann04}
{Brinchmann}, J., {Charlot}, S., {White}, S.~D.~M., {et~al.} 2004, \mnras, 351, 1151

\bibitem[{{Cedr{\'e}s} {et~al.}(2013){Cedr{\'e}s}, {Cepa}, {Bongiovanni}, {Casta{\~n}eda}, {S{\'a}nchez-Portal}, \& {Tomita}}]{cedres13}
{Cedr{\'e}s}, B., {Cepa}, J., {Bongiovanni}, {\'A}., {et~al.} 2013, \aap, 560, A59

\bibitem[{{Cepa} \& {Beckman}(1990)}]{cepa90}
{Cepa}, J. \& {Beckman}, J.~E. 1990, \apj, 349, 497

\bibitem[{{Chen} {et~al.}(2024){Chen}, {Grasha}, {Battisti}, {Wisnioski}, {Li}, {Park}, {Groves}, {Torrey}, {Mendel}, {Madore}, {Seibert}, {Sextl}, {Garcia}, {Rich}, {Beaton}, \& {Kewley}}]{chen24}
{Chen}, Q.-H., {Grasha}, K., {Battisti}, A.~J., {et~al.} 2024, \mnras, 534, 883

\bibitem[{{Chevance} {et~al.}(2020){Chevance}, {Kruijssen}, {Hygate}, {Schruba}, {Longmore}, {Groves}, {Henshaw}, {Herrera}, {Hughes}, {Jeffreson}, {Lang}, {Leroy}, {Meidt}, {Pety}, {Razza}, {Rosolowsky}, {Schinnerer}, {Bigiel}, {Blanc}, {Emsellem}, {Faesi}, {Glover}, {Haydon}, {Ho}, {Kreckel}, {Lee}, {Liu}, {Querejeta}, {Saito}, {Sun}, {Usero}, \& {Utomo}}]{chevance20}
{Chevance}, M., {Kruijssen}, J.~M.~D., {Hygate}, A. P.~S., {et~al.} 2020, \mnras, 493, 2872

\bibitem[{{Chevance} {et~al.}(2022){Chevance}, {Kruijssen}, {Krumholz}, {Groves}, {Keller}, {Hughes}, {Glover}, {Henshaw}, {Herrera}, {Kim}, {Leroy}, {Pety}, {Razza}, {Rosolowsky}, {Schinnerer}, {Schruba}, {Barnes}, {Bigiel}, {Blanc}, {Dale}, {Emsellem}, {Faesi}, {Grasha}, {Klessen}, {Kreckel}, {Liu}, {Longmore}, {Meidt}, {Querejeta}, {Saito}, {Sun}, \& {Usero}}]{chevance22}
{Chevance}, M., {Kruijssen}, J.~M.~D., {Krumholz}, M.~R., {et~al.} 2022, \mnras, 509, 272

\bibitem[{{Chevance} {et~al.}(2025){Chevance}, {Kruijssen}, \& {Longmore}}]{cool_whitepaper}
{Chevance}, M., {Kruijssen}, J.~M.~D., \& {Longmore}, S.~N. 2025, arXiv e-prints, arXiv:2501.13160

\bibitem[{{Chevance} {et~al.}(2023){Chevance}, {Krumholz}, {McLeod}, {Ostriker}, {Rosolowsky}, \& {Sternberg}}]{chevance23}
{Chevance}, M., {Krumholz}, M.~R., {McLeod}, A.~F., {et~al.} 2023, in Astronomical Society of the Pacific Conference Series, Vol. 534, Protostars and Planets VII, ed. S.~{Inutsuka}, Y.~{Aikawa}, T.~{Muto}, K.~{Tomida}, \& M.~{Tamura}, 1

\bibitem[{{Colombo} {et~al.}(2014){Colombo}, {Meidt}, {Schinnerer}, {Garc{\'\i}a-Burillo}, {Hughes}, {Pety}, {Leroy}, {Dobbs}, {Dumas}, {Thompson}, {Schuster}, \& {Kramer}}]{colombo14}
{Colombo}, D., {Meidt}, S.~E., {Schinnerer}, E., {et~al.} 2014, \apj, 784, 4

\bibitem[{{Corbelli} {et~al.}(2017){Corbelli}, {Braine}, {Bandiera}, {Brouillet}, {Combes}, {Druard}, {Gratier}, {Mata}, {Schuster}, {Xilouris}, \& {Palla}}]{Corbelli17}
{Corbelli}, E., {Braine}, J., {Bandiera}, R., {et~al.} 2017, \aap, 601, A146

\bibitem[{{Davis} {et~al.}(2014){Davis}, {Young}, {Crocker}, {Bureau}, {Blitz}, {Alatalo}, {Emsellem}, {Naab}, {Bayet}, {Bois}, {Bournaud}, {Cappellari}, {Davies}, {de Zeeuw}, {Duc}, {Khochfar}, {Krajnovi{\'c}}, {Kuntschner}, {McDermid}, {Morganti}, {Oosterloo}, {Sarzi}, {Scott}, {Serra}, \& {Weijmans}}]{davis14}
{Davis}, T.~A., {Young}, L.~M., {Crocker}, A.~F., {et~al.} 2014, \mnras, 444, 3427

\bibitem[{{den Brok} {et~al.}(2022){den Brok}, {Bigiel}, {Sliwa}, {Saito}, {Usero}, {Schinnerer}, {Leroy}, {Jim{\'e}nez-Donaire}, {Rosolowsky}, {Barnes}, {Puschnig}, {Pety}, {Schruba}, {Be{\v{s}}li{\'c}}, {Cao}, {Eibensteiner}, {Glover}, {Klessen}, {Kruijssen}, {Meidt}, {Neumann}, {Tomi{\v{c}}i{\'c}}, {Pan}, {Querejeta}, {Watkins}, {Williams}, \& {Wilner}}]{denBrok22}
{den Brok}, J.~S., {Bigiel}, F., {Sliwa}, K., {et~al.} 2022, \aap, 662, A89

\bibitem[{{den Brok} {et~al.}(2021){den Brok}, {Chatzigiannakis}, {Bigiel}, {Puschnig}, {Barnes}, {Leroy}, {Jim{\'e}nez-Donaire}, {Usero}, {Schinnerer}, {Rosolowsky}, {Faesi}, {Grasha}, {Hughes}, {Kruijssen}, {Liu}, {Neumann}, {Pety}, {Querejeta}, {Saito}, {Schruba}, \& {Stuber}}]{denbrok21}
{den Brok}, J.~S., {Chatzigiannakis}, D., {Bigiel}, F., {et~al.} 2021, \mnras, 504, 3221

\bibitem[{{Dobbs} \& {Baba}(2014)}]{dobbs14}
{Dobbs}, C. \& {Baba}, J. 2014, \pasa, 31, e035

\bibitem[{{Dobbs}(2011)}]{Dobbs2011}
{Dobbs}, C.~L. 2011, in EAS Publications Series, Vol.~52, EAS Publications Series, ed. M.~{R{\"o}llig}, R.~{Simon}, V.~{Ossenkopf}, \& J.~{Stutzki}, 87--93

\bibitem[{{Donovan Meyer} {et~al.}(2013){Donovan Meyer}, {Koda}, {Momose}, {Mooney}, {Egusa}, {Carty}, {Kennicutt}, {Kuno}, {Rebolledo}, {Sawada}, {Scoville}, \& {Wong}}]{donovan13}
{Donovan Meyer}, J., {Koda}, J., {Momose}, R., {et~al.} 2013, \apj, 772, 107

\bibitem[{{Downes} \& {Solomon}(1998)}]{downes98}
{Downes}, D. \& {Solomon}, P.~M. 1998, \apj, 507, 615

\bibitem[{{Egusa} {et~al.}(2009){Egusa}, {Kohno}, {Sofue}, {Nakanishi}, \& {Komugi}}]{egusa09}
{Egusa}, F., {Kohno}, K., {Sofue}, Y., {Nakanishi}, H., \& {Komugi}, S. 2009, \apj, 697, 1870

\bibitem[{{Egusa} {et~al.}(2004){Egusa}, {Sofue}, \& {Nakanishi}}]{egusa04}
{Egusa}, F., {Sofue}, Y., \& {Nakanishi}, H. 2004, \pasj, 56, L45

\bibitem[{{Elmegreen} \& {Elmegreen}(1986)}]{elmegreen86}
{Elmegreen}, B.~G. \& {Elmegreen}, D.~M. 1986, \apj, 311, 554

\bibitem[{{Emsellem} {et~al.}(2022){Emsellem}, {Schinnerer}, {Santoro}, {Belfiore}, {Pessa}, {McElroy}, {Blanc}, {Congiu}, {Groves}, {Ho}, {Kreckel}, {Razza}, {Sanchez-Blazquez}, {Egorov}, {Faesi}, {Klessen}, {Leroy}, {Meidt}, {Querejeta}, {Rosolowsky}, {Scheuermann}, {Anand}, {Barnes}, {Be{\v{s}}li{\'c}}, {Bigiel}, {Boquien}, {Cao}, {Chevance}, {Dale}, {Eibensteiner}, {Glover}, {Grasha}, {Henshaw}, {Hughes}, {Koch}, {Kruijssen}, {Lee}, {Liu}, {Pan}, {Pety}, {Saito}, {Sandstrom}, {Schruba}, {Sun}, {Thilker}, {Usero}, {Watkins}, \& {Williams}}]{emsellem22}
{Emsellem}, E., {Schinnerer}, E., {Santoro}, F., {et~al.} 2022, \aap, 659, A191

\bibitem[{{Engargiola} {et~al.}(2003){Engargiola}, {Plambeck}, {Rosolowsky}, \& {Blitz}}]{Engargiola2003}
{Engargiola}, G., {Plambeck}, R.~L., {Rosolowsky}, E., \& {Blitz}, L. 2003, \apjs, 149, 343

\bibitem[{{Feldmann} {et~al.}(2011){Feldmann}, {Gnedin}, \& {Kravtsov}}]{feldmann11}
{Feldmann}, R., {Gnedin}, N.~Y., \& {Kravtsov}, A.~V. 2011, \apj, 732, 115

\bibitem[{{Fitzpatrick}(1999)}]{fitzpatrick99}
{Fitzpatrick}, E.~L. 1999, \pasp, 111, 63

\bibitem[{{Foyle} {et~al.}(2010){Foyle}, {Rix}, {Walter}, \& {Leroy}}]{foyle10}
{Foyle}, K., {Rix}, H.~W., {Walter}, F., \& {Leroy}, A.~K. 2010, \apj, 725, 534

\bibitem[{{Fukui} {et~al.}(2014){Fukui}, {Ohama}, {Hanaoka}, {Furukawa}, {Torii}, {Dawson}, {Mizuno}, {Hasegawa}, {Fukuda}, {Soga}, {Moribe}, {Kuroda}, {Hayakawa}, {Kawamura}, {Kuwahara}, {Yamamoto}, {Okuda}, {Onishi}, {Maezawa}, \& {Mizuno}}]{fukui14}
{Fukui}, Y., {Ohama}, A., {Hanaoka}, N., {et~al.} 2014, \apj, 780, 36

\bibitem[{{Garcia-Burillo} {et~al.}(1993){Garcia-Burillo}, {Guelin}, \& {Cernicharo}}]{GarciaBurillo1993}
{Garcia-Burillo}, S., {Guelin}, M., \& {Cernicharo}, J. 1993, \aap, 274, 123

\bibitem[{{Gensior} {et~al.}(2020){Gensior}, {Kruijssen}, \& {Keller}}]{gensior20}
{Gensior}, J., {Kruijssen}, J.~M.~D., \& {Keller}, B.~W. 2020, \mnras, 495, 199

\bibitem[{{Grasha} {et~al.}(2019){Grasha}, {Calzetti}, {Adamo}, {Kennicutt}, {Elmegreen}, {Messa}, {Dale}, {Fedorenko}, {Mahadevan}, {Grebel}, {Fumagalli}, {Kim}, {Dobbs}, {Gouliermis}, {Ashworth}, {Gallagher}, {Smith}, {Tosi}, {Whitmore}, {Schinnerer}, {Colombo}, {Hughes}, {Leroy}, \& {Meidt}}]{grasha19}
{Grasha}, K., {Calzetti}, D., {Adamo}, A., {et~al.} 2019, \mnras, 483, 4707

\bibitem[{{Haydon} {et~al.}(2020{\natexlab{a}}){Haydon}, {Fujimoto}, {Chevance}, {Kruijssen}, {Krumholz}, \& {Longmore}}]{haydon20_ext}
{Haydon}, D.~T., {Fujimoto}, Y., {Chevance}, M., {et~al.} 2020{\natexlab{a}}, \mnras, 497, 5076

\bibitem[{{Haydon} {et~al.}(2020{\natexlab{b}}){Haydon}, {Kruijssen}, {Chevance}, {Hygate}, {Krumholz}, {Schruba}, \& {Longmore}}]{haydon20}
{Haydon}, D.~T., {Kruijssen}, J.~M.~D., {Chevance}, M., {et~al.} 2020{\natexlab{b}}, \mnras, 498, 235

\bibitem[{{Henry} {et~al.}(2003){Henry}, {Quillen}, \& {Gutermuth}}]{henry03}
{Henry}, A.~L., {Quillen}, A.~C., \& {Gutermuth}, R. 2003, \aj, 126, 2831

\bibitem[{{Hirota} {et~al.}(2018){Hirota}, {Egusa}, {Baba}, {Kuno}, {Muraoka}, {Tosaki}, {Miura}, {Nakanishi}, \& {Kawabe}}]{hirota18}
{Hirota}, A., {Egusa}, F., {Baba}, J., {et~al.} 2018, \pasj, 70, 73

\bibitem[{{Hitschfeld} {et~al.}(2009){Hitschfeld}, {Kramer}, {Schuster}, {Garcia-Burillo}, \& {Stutzki}}]{Hitschfeld2009}
{Hitschfeld}, M., {Kramer}, C., {Schuster}, K.~F., {Garcia-Burillo}, S., \& {Stutzki}, J. 2009, \aap, 495, 795

\bibitem[{{Hughes} {et~al.}(2013){Hughes}, {Meidt}, {Colombo}, {Schinnerer}, {Pety}, {Leroy}, {Dobbs}, {Garc{\'\i}a-Burillo}, {Thompson}, {Dumas}, {Schuster}, \& {Kramer}}]{hughes13}
{Hughes}, A., {Meidt}, S.~E., {Colombo}, D., {et~al.} 2013, \apj, 779, 46

\bibitem[{{Hygate} {et~al.}(2019){Hygate}, {Kruijssen}, {Chevance}, {Schruba}, {Haydon}, \& {Longmore}}]{hygate19}
{Hygate}, A. P.~S., {Kruijssen}, J.~M.~D., {Chevance}, M., {et~al.} 2019, \mnras, 488, 2800

\bibitem[{{Jeffreson} \& {Kruijssen}(2018)}]{jeffreson18}
{Jeffreson}, S. M.~R. \& {Kruijssen}, J.~M.~D. 2018, \mnras, 476, 3688

\bibitem[{{Kawamura} {et~al.}(2009){Kawamura}, {Mizuno}, {Minamidani}, {Filipovi{\'c}}, {Staveley-Smith}, {Kim}, {Mizuno}, {Onishi}, {Mizuno}, \& {Fukui}}]{kawamura09}
{Kawamura}, A., {Mizuno}, Y., {Minamidani}, T., {et~al.} 2009, \apjs, 184, 1

\bibitem[{{Kennicutt}(1998)}]{kennicutt98}
{Kennicutt}, Robert~C., J. 1998, \apj, 498, 541

\bibitem[{{Kennicutt} \& {Evans}(2012)}]{kennicutt12}
{Kennicutt}, R.~C. \& {Evans}, N.~J. 2012, \araa, 50, 531

\bibitem[{{Kim} {et~al.}(2023){Kim}, {Chevance}, {Kruijssen}, {Barnes}, {Bigiel}, {Blanc}, {Boquien}, {Cao}, {Congiu}, {Dale}, {Egorov}, {Faesi}, {Glover}, {Grasha}, {Groves}, {Hassani}, {Hughes}, {Klessen}, {Kreckel}, {Larson}, {Lee}, {Leroy}, {Liu}, {Longmore}, {Meidt}, {Pan}, {Pety}, {Querejeta}, {Rosolowsky}, {Saito}, {Sandstrom}, {Schinnerer}, {Smith}, {Usero}, {Watkins}, \& {Williams}}]{kim23}
{Kim}, J., {Chevance}, M., {Kruijssen}, J.~M.~D., {et~al.} 2023, \apjl, 944, L20

\bibitem[{{Kim} {et~al.}(2022){Kim}, {Chevance}, {Kruijssen}, {Leroy}, {Schruba}, {Barnes}, {Bigiel}, {Blanc}, {Cao}, {Congiu}, {Dale}, {Faesi}, {Glover}, {Grasha}, {Groves}, {Hughes}, {Klessen}, {Kreckel}, {McElroy}, {Pan}, {Pety}, {Querejeta}, {Razza}, {Rosolowsky}, {Saito}, {Schinnerer}, {Sun}, {Tomi{\v{c}}i{\'c}}, {Usero}, \& {Williams}}]{kim22}
{Kim}, J., {Chevance}, M., {Kruijssen}, J.~M.~D., {et~al.} 2022, \mnras, 516, 3006

\bibitem[{{Kim} {et~al.}(2021){Kim}, {Chevance}, {Kruijssen}, {Schruba}, {Sandstrom}, {Barnes}, {Bigiel}, {Blanc}, {Cao}, {Dale}, {Faesi}, {Glover}, {Grasha}, {Groves}, {Herrera}, {Klessen}, {Kreckel}, {Lee}, {Leroy}, {Pety}, {Querejeta}, {Schinnerer}, {Sun}, {Usero}, {Ward}, \& {Williams}}]{kim21}
{Kim}, J., {Chevance}, M., {Kruijssen}, J.~M.~D., {et~al.} 2021, \mnras, 504, 487

\bibitem[{{Knapen} \& {Beckman}(1996)}]{knapen96}
{Knapen}, J.~H. \& {Beckman}, J.~E. 1996, \mnras, 283, 251

\bibitem[{{Koda}(2013)}]{Koda2013}
{Koda}, J. 2013, in Astronomical Society of the Pacific Conference Series, Vol. 476, New Trends in Radio Astronomy in the ALMA Era: The 30th Anniversary of Nobeyama Radio Observatory, ed. R.~{Kawabe}, N.~{Kuno}, \& S.~{Yamamoto}, 49

\bibitem[{{Koda} {et~al.}(2012){Koda}, {Scoville}, {Hasegawa}, {Calzetti}, {Donovan Meyer}, {Egusa}, {Kennicutt}, {Kuno}, {Louie}, {Momose}, {Sawada}, {Sorai}, \& {Umei}}]{Koda2012}
{Koda}, J., {Scoville}, N., {Hasegawa}, T., {et~al.} 2012, \apj, 761, 41

\bibitem[{{Koda} {et~al.}(2009){Koda}, {Scoville}, {Sawada}, {La Vigne}, {Vogel}, {Potts}, {Carpenter}, {Corder}, {Wright}, {White}, {Zauderer}, {Patience}, {Sargent}, {Bock}, {Hawkins}, {Hodges}, {Kemball}, {Lamb}, {Plambeck}, {Pound}, {Scott}, {Teuben}, \& {Woody}}]{Koda2009}
{Koda}, J., {Scoville}, N., {Sawada}, T., {et~al.} 2009, \apjl, 700, L132

\bibitem[{{Kreckel} {et~al.}(2016){Kreckel}, {Blanc}, {Schinnerer}, {Groves}, {Adamo}, {Hughes}, \& {Meidt}}]{kreckel16}
{Kreckel}, K., {Blanc}, G.~A., {Schinnerer}, E., {et~al.} 2016, \apj, 827, 103

\bibitem[{{Kruijssen}(2014)}]{kruijssen14c}
{Kruijssen}, J.~M.~D. 2014, Classical and Quantum Gravity, 31, 244006

\bibitem[{{Kruijssen} {et~al.}(2024){Kruijssen}, {Chevance}, {Longmore}, {Ginsburg}, {Ramambason}, \& {Romanelli}}]{kruijssen24}
{Kruijssen}, J.~M.~D., {Chevance}, M., {Longmore}, S.~N., {et~al.} 2024, OJA~submitted, arXiv:2404.14495

\bibitem[{{Kruijssen} \& {Longmore}(2014)}]{kruijssen14}
{Kruijssen}, J.~M.~D. \& {Longmore}, S.~N. 2014, \mnras, 439, 3239

\bibitem[{{Kruijssen} {et~al.}(2019){Kruijssen}, {Schruba}, {Chevance}, {Longmore}, {Hygate}, {Haydon}, {McLeod}, {Dalcanton}, {Tacconi}, \& {van Dishoeck}}]{kruijssen19}
{Kruijssen}, J.~M.~D., {Schruba}, A., {Chevance}, M., {et~al.} 2019, \nat, 569, 519

\bibitem[{{Kruijssen} {et~al.}(2018){Kruijssen}, {Schruba}, {Hygate}, {Hu}, {Haydon}, \& {Longmore}}]{kruijssen18}
{Kruijssen}, J.~M.~D., {Schruba}, A., {Hygate}, A. P.~S., {et~al.} 2018, \mnras, 479, 1866

\bibitem[{{Krumholz} \& {McKee}(2005)}]{krumholz05}
{Krumholz}, M.~R. \& {McKee}, C.~F. 2005, \apj, 630, 250

\bibitem[{{Kuno} {et~al.}(2007){Kuno}, {Sato}, {Nakanishi}, {Hirota}, {Tosaki}, {Shioya}, {Sorai}, {Nakai}, {Nishiyama}, \& {Vila-Vilar{\'o}}}]{Kuno07}
{Kuno}, N., {Sato}, N., {Nakanishi}, H., {et~al.} 2007, \pasj, 59, 117

\bibitem[{{Lang} {et~al.}(2020){Lang}, {Meidt}, {Rosolowsky}, {Nofech}, {Schinnerer}, {Leroy}, {Emsellem}, {Pessa}, {Glover}, {Groves}, {Hughes}, {Kruijssen}, {Querejeta}, {Schruba}, {Bigiel}, {Blanc}, {Chevance}, {Colombo}, {Faesi}, {Henshaw}, {Herrera}, {Liu}, {Pety}, {Puschnig}, {Saito}, {Sun}, \& {Usero}}]{lang20}
{Lang}, P., {Meidt}, S.~E., {Rosolowsky}, E., {et~al.} 2020, \apj, 897, 122

\bibitem[{{Leroy} {et~al.}(2021{\natexlab{a}}){Leroy}, {Hughes}, {Liu}, {Pety}, {Rosolowsky}, {Saito}, {Schinnerer}, {Schruba}, {Usero}, {Faesi}, {Herrera}, {Chevance}, {Hygate}, {Kepley}, {Koch}, {Querejeta}, {Sliwa}, {Will}, {Wilson}, {Anand}, {Barnes}, {Belfiore}, {Be{\v{s}}li{\'c}}, {Bigiel}, {Blanc}, {Bolatto}, {Boquien}, {Cao}, {Chandar}, {Chastenet}, {Chiang}, {Congiu}, {Dale}, {Deger}, {den Brok}, {Eibensteiner}, {Emsellem}, {Garc{\'\i}a-Rodr{\'\i}guez}, {Glover}, {Grasha}, {Groves}, {Henshaw}, {Jim{\'e}nez Donaire}, {Kim}, {Klessen}, {Kreckel}, {Kruijssen}, {Larson}, {Lee}, {Mayker}, {McElroy}, {Meidt}, {Mok}, {Pan}, {Puschnig}, {Razza}, {S{\'a}nchez-Bl'azquez}, {Sandstrom}, {Santoro}, {Sardone}, {Scheuermann}, {Sun}, {Thilker}, {Turner}, {Ubeda}, {Utomo}, {Watkins}, \& {Williams}}]{leroy21_pipe}
{Leroy}, A.~K., {Hughes}, A., {Liu}, D., {et~al.} 2021{\natexlab{a}}, \apjs, 255, 19

\bibitem[{{Leroy} {et~al.}(2019){Leroy}, {Sandstrom}, {Lang}, {Lewis}, {Salim}, {Behrens}, {Chastenet}, {Chiang}, {Gallagher}, {Kessler}, \& {Utomo}}]{leroy19}
{Leroy}, A.~K., {Sandstrom}, K.~M., {Lang}, D., {et~al.} 2019, \apjs, 244, 24

\bibitem[{{Leroy} {et~al.}(2017){Leroy}, {Schinnerer}, {Hughes}, {Kruijssen}, {Meidt}, {Schruba}, {Sun}, {Bigiel}, {Aniano}, {Blanc}, {Bolatto}, {Chevance}, {Colombo}, {Gallagher}, {Garcia-Burillo}, {Kramer}, {Querejeta}, {Pety}, {Thompson}, \& {Usero}}]{leroy17}
{Leroy}, A.~K., {Schinnerer}, E., {Hughes}, A., {et~al.} 2017, \apj, 846, 71

\bibitem[{{Leroy} {et~al.}(2021{\natexlab{b}}){Leroy}, {Schinnerer}, {Hughes}, {Rosolowsky}, {Pety}, {Schruba}, {Usero}, {Blanc}, {Chevance}, {Emsellem}, {Faesi}, {Herrera}, {Liu}, {Meidt}, {Querejeta}, {Saito}, {Sandstrom}, {Sun}, {Williams}, {Anand}, {Barnes}, {Behrens}, {Belfiore}, {Benincasa}, {Be{\v{s}}li{\'c}}, {Bigiel}, {Bolatto}, {den Brok}, {Cao}, {Chandar}, {Chastenet}, {Chiang}, {Congiu}, {Dale}, {Deger}, {Eibensteiner}, {Egorov}, {Garc{\'\i}a-Rodr{\'\i}guez}, {Glover}, {Grasha}, {Henshaw}, {Ho}, {Kepley}, {Kim}, {Klessen}, {Kreckel}, {Koch}, {Kruijssen}, {Larson}, {Lee}, {Lopez}, {Machado}, {Mayker}, {McElroy}, {Murphy}, {Ostriker}, {Pan}, {Pessa}, {Puschnig}, {Razza}, {S{\'a}nchez-Bl{\'a}zquez}, {Santoro}, {Sardone}, {Scheuermann}, {Sliwa}, {Sormani}, {Stuber}, {Thilker}, {Turner}, {Utomo}, {Watkins}, \& {Whitmore}}]{Leroy2021a}
{Leroy}, A.~K., {Schinnerer}, E., {Hughes}, A., {et~al.} 2021{\natexlab{b}}, \apjs, 257, 43

\bibitem[{{Leroy} {et~al.}(2025){Leroy}, {Sun}, {Meidt}, {Agertz}, {Chiang}, {Gensior}, {Glover}, {Gnedin}, {Hughes}, {Schinnerer}, {Barnes}, {Bigiel}, {Bolatto}, {Colombo}, {den Brok}, {Chevance}, {Chown}, {Eibensteiner}, {Gleis}, {Grasha}, {Henshaw}, {Klessen}, {Koch}, {Oakes}, {Pan}, {Querejeta}, {Rosolowsky}, {Saito}, {Sandstrom}, {Sarbadhicary}, {Teng}, {Usero}, {Utomo}, \& {Williams}}]{Leroy2025}
{Leroy}, A.~K., {Sun}, J., {Meidt}, S., {et~al.} 2025, arXiv e-prints, arXiv:2502.04481

\bibitem[{{Leroy} {et~al.}(2013){Leroy}, {Walter}, {Sandstrom}, {Schruba}, {Munoz-Mateos}, {Bigiel}, {Bolatto}, {Brinks}, {de Blok}, {Meidt}, {Rix}, {Rosolowsky}, {Schinnerer}, {Schuster}, \& {Usero}}]{leroy13}
{Leroy}, A.~K., {Walter}, F., {Sandstrom}, K., {et~al.} 2013, \aj, 146, 19

\bibitem[{{Liu} {et~al.}(2021){Liu}, {Bureau}, {Blitz}, {Davis}, {Onishi}, {Smith}, {North}, \& {Iguchi}}]{Liu2021}
{Liu}, L., {Bureau}, M., {Blitz}, L., {et~al.} 2021, \mnras, 505, 4048

\bibitem[{{Lord} \& {Young}(1990)}]{lord90}
{Lord}, S.~D. \& {Young}, J.~S. 1990, \apj, 356, 135

\bibitem[{{Lu} {et~al.}(2022){Lu}, {Boyce}, {Haggard}, {Bureau}, {Liang}, {Liu}, {Choi}, {Cappellari}, {Chemin}, {Chevance}, {Davis}, {Drissen}, {Elford}, {Gensior}, {Kruijssen}, {Martin}, {Mass{\'e}}, {Robert}, {Ruffa}, {Rousseau-Nepton}, {Sarzi}, {Savard}, \& {Williams}}]{lu22}
{Lu}, A., {Boyce}, H., {Haggard}, D., {et~al.} 2022, \mnras, 514, 5035

\bibitem[{{Martig} {et~al.}(2009){Martig}, {Bournaud}, {Teyssier}, \& {Dekel}}]{Martig2009}
{Martig}, M., {Bournaud}, F., {Teyssier}, R., \& {Dekel}, A. 2009, \apj, 707, 250

\bibitem[{{Mathis}(1986)}]{Mathis1986}
{Mathis}, J.~S. 1986, \apj, 301, 423

\bibitem[{{Meidt} {et~al.}(2015){Meidt}, {Hughes}, {Dobbs}, {Pety}, {Thompson}, {Garc{\'\i}a-Burillo}, {Leroy}, {Schinnerer}, {Colombo}, {Querejeta}, {Kramer}, {Schuster}, \& {Dumas}}]{Meidt2015}
{Meidt}, S.~E., {Hughes}, A., {Dobbs}, C.~L., {et~al.} 2015, \apj, 806, 72

\bibitem[{{Meidt} {et~al.}(2021){Meidt}, {Leroy}, {Querejeta}, {Schinnerer}, {Sun}, {van der Wel}, {Emsellem}, {Henshaw}, {Hughes}, {Kruijssen}, {Rosolowsky}, {Schruba}, {Barnes}, {Bigiel}, {Blanc}, {Chevance}, {Cao}, {Dale}, {Faesi}, {Glover}, {Grasha}, {Groves}, {Herrera}, {Klessen}, {Kreckel}, {Liu}, {Pan}, {Pety}, {Saito}, {Usero}, {Watkins}, \& {Williams}}]{Meidt21}
{Meidt}, S.~E., {Leroy}, A.~K., {Querejeta}, M., {et~al.} 2021, \apj, 913, 113

\bibitem[{{Meidt} {et~al.}(2018){Meidt}, {Leroy}, {Rosolowsky}, {Kruijssen}, {Schinnerer}, {Schruba}, {Pety}, {Blanc}, {Bigiel}, {Chevance}, {Hughes}, {Querejeta}, \& {Usero}}]{meidt18}
{Meidt}, S.~E., {Leroy}, A.~K., {Rosolowsky}, E., {et~al.} 2018, \apj, 854, 100

\bibitem[{{Meidt} {et~al.}(2013){Meidt}, {Schinnerer}, {Garc{\'\i}a-Burillo}, {Hughes}, {Colombo}, {Pety}, {Dobbs}, {Schuster}, {Kramer}, {Leroy}, {Dumas}, \& {Thompson}}]{meidt13}
{Meidt}, S.~E., {Schinnerer}, E., {Garc{\'\i}a-Burillo}, S., {et~al.} 2013, \apj, 779, 45

\bibitem[{{Nakanishi} \& {Sofue}(2003)}]{NakanishiSofue2003}
{Nakanishi}, H. \& {Sofue}, Y. 2003, \pasj, 55, 191

\bibitem[{{Onodera} {et~al.}(2010){Onodera}, {Kuno}, {Tosaki}, {Kohno}, {Nakanishi}, {Sawada}, {Muraoka}, {Komugi}, {Miura}, {Kaneko}, {Hirota}, \& {Kawabe}}]{onodera10}
{Onodera}, S., {Kuno}, N., {Tosaki}, T., {et~al.} 2010, \apjl, 722, L127

\bibitem[{{Padoan} {et~al.}(2012){Padoan}, {Haugb{\o}lle}, \& {Nordlund}}]{padoan12}
{Padoan}, P., {Haugb{\o}lle}, T., \& {Nordlund}, {\r{A}}. 2012, \apjl, 759, L27

\bibitem[{{Pilyugin} \& {Grebel}(2016)}]{pilyugin16}
{Pilyugin}, L.~S. \& {Grebel}, E.~K. 2016, \mnras, 457, 3678

\bibitem[{{Pilyugin} {et~al.}(2014){Pilyugin}, {Grebel}, {Zinchenko}, \& {Kniazev}}]{pilyugin14}
{Pilyugin}, L.~S., {Grebel}, E.~K., {Zinchenko}, I.~A., \& {Kniazev}, A.~Y. 2014, \aj, 148, 134

\bibitem[{{Querejeta} {et~al.}(2024){Querejeta}, {Leroy}, {Meidt}, {Schinnerer}, {Belfiore}, {Emsellem}, {Klessen}, {Sun}, {Sormani}, {Be{\v{s}}li{\'c}}, {Cao}, {Chevance}, {Colombo}, {Dale}, {Garc{\'\i}a-Burillo}, {Glover}, {Grasha}, {Groves}, {Koch}, {Neumann}, {Pan}, {Pessa}, {Pety}, {Pinna}, {Ramambason}, {Razza}, {Romanelli}, {Rosolowsky}, {Ruiz-Garc{\'\i}a}, {S{\'a}nchez-Bl{\'a}zquez}, {Smith}, {Stuber}, {Ubeda}, {Usero}, \& {Williams}}]{querejeta24}
{Querejeta}, M., {Leroy}, A.~K., {Meidt}, S.~E., {et~al.} 2024, \aap, 687, A293

\bibitem[{{Querejeta} {et~al.}(2021){Querejeta}, {Schinnerer}, {Meidt}, {Sun}, {Leroy}, {Emsellem}, {Klessen}, {Mu{\~n}oz-Mateos}, {Salo}, {Laurikainen}, {Be{\v{s}}li{\'c}}, {Blanc}, {Chevance}, {Dale}, {Eibensteiner}, {Faesi}, {Garc{\'\i}a-Rodr{\'\i}guez}, {Glover}, {Grasha}, {Henshaw}, {Herrera}, {Hughes}, {Kreckel}, {Kruijssen}, {Liu}, {Murphy}, {Pan}, {Pety}, {Razza}, {Rosolowsky}, {Saito}, {Schruba}, {Usero}, {Watkins}, \& {Williams}}]{querejeta21}
{Querejeta}, M., {Schinnerer}, E., {Meidt}, S., {et~al.} 2021, \aap, 656, A133

\bibitem[{{Querejeta} {et~al.}(2019){Querejeta}, {Schinnerer}, {Schruba}, {Murphy}, {Meidt}, {Usero}, {Leroy}, {Pety}, {Bigiel}, {Chevance}, {Faesi}, {Gallagher}, {Garc{\'\i}a-Burillo}, {Glover}, {Hygate}, {Jim{\'e}nez-Donaire}, {Kruijssen}, {Momjian}, {Rosolowsky}, \& {Utomo}}]{querejeta19}
{Querejeta}, M., {Schinnerer}, E., {Schruba}, A., {et~al.} 2019, \aap, 625, A19

\bibitem[{{Ragan} {et~al.}(2018){Ragan}, {Moore}, {Eden}, {Hoare}, {Urquhart}, {Elia}, \& {Molinari}}]{ragan18}
{Ragan}, S.~E., {Moore}, T.~J.~T., {Eden}, D.~J., {et~al.} 2018, \mnras, 479, 2361

\bibitem[{{Roberts}(1969)}]{roberts69}
{Roberts}, W.~W. 1969, \apj, 158, 123

\bibitem[{{Rosolowsky} \& {Blitz}(2005)}]{rosolowsky05}
{Rosolowsky}, E. \& {Blitz}, L. 2005, \apj, 623, 826

\bibitem[{{Rosolowsky} {et~al.}(2021){Rosolowsky}, {Hughes}, {Leroy}, {Sun}, {Querejeta}, {Schruba}, {Usero}, {Herrera}, {Liu}, {Pety}, {Saito}, {Be{\v{s}}li{\'c}}, {Bigiel}, {Blanc}, {Chevance}, {Dale}, {Deger}, {Faesi}, {Glover}, {Henshaw}, {Klessen}, {Kruijssen}, {Larson}, {Lee}, {Meidt}, {Mok}, {Schinnerer}, {Thilker}, \& {Williams}}]{rosolowsky21}
{Rosolowsky}, E., {Hughes}, A., {Leroy}, A.~K., {et~al.} 2021, \mnras, 502, 1218

\bibitem[{{Saintonge} {et~al.}(2017){Saintonge}, {Catinella}, {Tacconi}, {Kauffmann}, {Genzel}, {Cortese}, {Dav{\'e}}, {Fletcher}, {Graci{\'a}-Carpio}, {Kramer}, {Heckman}, {Janowiecki}, {Lutz}, {Rosario}, {Schiminovich}, {Schuster}, {Wang}, {Wuyts}, {Borthakur}, {Lamperti}, \& {Roberts-Borsani}}]{Saintonge17}
{Saintonge}, A., {Catinella}, B., {Tacconi}, L.~J., {et~al.} 2017, \apjs, 233, 22

\bibitem[{{S{\'a}nchez} {et~al.}(2019){S{\'a}nchez}, {Barrera-Ballesteros}, {L{\'o}pez-Cob{\'a}}, {Brough}, {Bryant}, {Bland-Hawthorn}, {Croom}, {van de Sande}, {Cortese}, {Goodwin}, {Lawrence}, {L{\'o}pez-S{\'a}nchez}, {Sweet}, {Owers}, {Richards}, \& {Walcher}}]{sanchez19}
{S{\'a}nchez}, S.~F., {Barrera-Ballesteros}, J.~K., {L{\'o}pez-Cob{\'a}}, C., {et~al.} 2019, \mnras, 484, 3042

\bibitem[{{S{\'a}nchez} {et~al.}(2014){S{\'a}nchez}, {Rosales-Ortega}, {Iglesias-P{\'a}ramo}, {Moll{\'a}}, {Barrera-Ballesteros}, {Marino}, {P{\'e}rez}, {S{\'a}nchez-Blazquez}, {Gonz{\'a}lez Delgado}, {Cid Fernandes}, {de Lorenzo-C{\'a}ceres}, {Mendez-Abreu}, {Galbany}, {Falcon-Barroso}, {Miralles-Caballero}, {Husemann}, {Garc{\'\i}a-Benito}, {Mast}, {Walcher}, {Gil de Paz}, {Garc{\'\i}a-Lorenzo}, {Jungwiert}, {V{\'\i}lchez}, {J{\'\i}lkov{\'a}}, {Lyubenova}, {Cortijo-Ferrero}, {D{\'\i}az}, {Wisotzki}, {M{\'a}rquez}, {Bland-Hawthorn}, {Ellis}, {van de Ven}, {Jahnke}, {Papaderos}, {Gomes}, {Mendoza}, \& {L{\'o}pez-S{\'a}nchez}}]{sanchez14}
{S{\'a}nchez}, S.~F., {Rosales-Ortega}, F.~F., {Iglesias-P{\'a}ramo}, J., {et~al.} 2014, \aap, 563, A49

\bibitem[{{Schinnerer} {et~al.}(2019){Schinnerer}, {Hughes}, {Leroy}, {Groves}, {Blanc}, {Kreckel}, {Bigiel}, {Chevance}, {Dale}, {Emsellem}, {Faesi}, {Glover}, {Grasha}, {Henshaw}, {Hygate}, {Kruijssen}, {Meidt}, {Pety}, {Querejeta}, {Rosolowsky}, {Saito}, {Schruba}, {Sun}, \& {Utomo}}]{schinnerer19}
{Schinnerer}, E., {Hughes}, A., {Leroy}, A., {et~al.} 2019, \apj, 887, 49

\bibitem[{{Schinnerer} \& {Leroy}(2024)}]{schinnerer&leroy24}
{Schinnerer}, E. \& {Leroy}, A.~K. 2024, \araa, 62, 369

\bibitem[{{Schinnerer} {et~al.}(2013){Schinnerer}, {Meidt}, {Pety}, {Hughes}, {Colombo}, {Garc{\'\i}a-Burillo}, {Schuster}, {Dumas}, {Dobbs}, {Leroy}, {Kramer}, {Thompson}, \& {Regan}}]{schinnerer13}
{Schinnerer}, E., {Meidt}, S.~E., {Pety}, J., {et~al.} 2013, \apj, 779, 42

\bibitem[{{Schmidt}(1959)}]{schmidt59}
{Schmidt}, M. 1959, \apj, 129, 243

\bibitem[{Scholz \& Stephens(1987)}]{scholz87}
Scholz, F.~W. \& Stephens, M.~A. 1987, Journal of the American Statistical Association, 82, 918

\bibitem[{{Schruba} {et~al.}(2010){Schruba}, {Leroy}, {Walter}, {Sandstrom}, \& {Rosolowsky}}]{schruba10}
{Schruba}, A., {Leroy}, A.~K., {Walter}, F., {Sandstrom}, K., \& {Rosolowsky}, E. 2010, \apj, 722, 1699

\bibitem[{{Scoville} \& {Hersh}(1979)}]{scoville79}
{Scoville}, N.~Z. \& {Hersh}, K. 1979, \apj, 229, 578

\bibitem[{{Scoville} \& {Wilson}(2004)}]{Scoville2004}
{Scoville}, N.~Z. \& {Wilson}, C.~D. 2004, in Astronomical Society of the Pacific Conference Series, Vol. 322, The Formation and Evolution of Massive Young Star Clusters, ed. H.~J.~G.~L.~M. {Lamers}, L.~J. {Smith}, \& A.~{Nota}, 245

\bibitem[{{Seigar} \& {James}(2002)}]{seigar02}
{Seigar}, M.~S. \& {James}, P.~A. 2002, \mnras, 337, 1113

\bibitem[{{Sembach} {et~al.}(2000){Sembach}, {Howk}, {Ryans}, \& {Keenan}}]{Sembach2000}
{Sembach}, K.~R., {Howk}, J.~C., {Ryans}, R. S.~I., \& {Keenan}, F.~P. 2000, \apj, 528, 310

\bibitem[{{Silk}(1997)}]{Silk97}
{Silk}, J. 1997, \apj, 481, 703

\bibitem[{{Sorai} {et~al.}(2019){Sorai}, {Kuno}, {Muraoka}, {Miyamoto}, {Kaneko}, {Nakanishi}, {Nakai}, {Yanagitani}, {Tanaka}, {Sato}, {Salak}, {Umei}, {Morokuma-Matsui}, {Matsumoto}, {Ueno}, {Pan}, {Noma}, {Takeuchi}, {Yoda}, {Kuroda}, {Yasuda}, {Yajima}, {Oi}, {Shibata}, {Seta}, {Watanabe}, {Kita}, {Komatsuzaki}, {Kajikawa}, {Yashima}, {Cooray}, {Baji}, {Segawa}, {Tashiro}, {Takeda}, {Kishida}, {Hatakeyama}, {Tomiyasu}, \& {Saita}}]{Sorai19}
{Sorai}, K., {Kuno}, N., {Muraoka}, K., {et~al.} 2019, \pasj, 71, S14

\bibitem[{{Sun} {et~al.}(2024){Sun}, {Calzetti}, \& {Battisti}}]{sun&calzetti24}
{Sun}, B., {Calzetti}, D., \& {Battisti}, A.~J. 2024, \apj, 973, 137

\bibitem[{{Sun} {et~al.}(2022){Sun}, {Leroy}, {Rosolowsky}, {Hughes}, {Schinnerer}, {Schruba}, {Koch}, {Blanc}, {Chiang}, {Groves}, {Liu}, {Meidt}, {Pan}, {Pety}, {Querejeta}, {Saito}, {Sandstrom}, {Sardone}, {Usero}, {Utomo}, {Williams}, {Barnes}, {Benincasa}, {Bigiel}, {Bolatto}, {Boquien}, {Chevance}, {Dale}, {Deger}, {Emsellem}, {Glover}, {Grasha}, {Henshaw}, {Klessen}, {Kreckel}, {Kruijssen}, {Ostriker}, \& {Thilker}}]{sun22}
{Sun}, J., {Leroy}, A.~K., {Rosolowsky}, E., {et~al.} 2022, \aj, 164, 43

\bibitem[{{Sun} {et~al.}(2020){Sun}, {Leroy}, {Schinnerer}, {Hughes}, {Rosolowsky}, {Querejeta}, {Schruba}, {Liu}, {Saito}, {Herrera}, {Faesi}, {Usero}, {Pety}, {Kruijssen}, {Ostriker}, {Bigiel}, {Blanc}, {Bolatto}, {Boquien}, {Chevance}, {Dale}, {Deger}, {Emsellem}, {Glover}, {Grasha}, {Groves}, {Henshaw}, {Jimenez-Donaire}, {Kim}, {Klessen}, {Kreckel}, {Lee}, {Meidt}, {Sandstrom}, {Sardone}, {Utomo}, \& {Williams}}]{sun20}
{Sun}, J., {Leroy}, A.~K., {Schinnerer}, E., {et~al.} 2020, \apjl, 901, L8

\bibitem[{{Vogel} {et~al.}(1988){Vogel}, {Kulkarni}, \& {Scoville}}]{Vogel1988}
{Vogel}, S.~N., {Kulkarni}, S.~R., \& {Scoville}, N.~Z. 1988, \nat, 334, 402

\bibitem[{{Ward} {et~al.}(2020){Ward}, {Chevance}, {Kruijssen}, {Hygate}, {Schruba}, \& {Longmore}}]{ward20_HI}
{Ward}, J.~L., {Chevance}, M., {Kruijssen}, J.~M.~D., {et~al.} 2020, \mnras, 497, 2286

\bibitem[{{Ward} {et~al.}(2022){Ward}, {Kruijssen}, {Chevance}, {Kim}, \& {Longmore}}]{ward22}
{Ward}, J.~L., {Kruijssen}, J.~M.~D., {Chevance}, M., {Kim}, J., \& {Longmore}, S.~N. 2022, \mnras, 516, 4025

\bibitem[{Wilcoxon(1945)}]{Wilcoxon45}
Wilcoxon, F. 1945, Biometrics Bulletin, 1, 80

\bibitem[{{Williams} {et~al.}(1994){Williams}, {de Geus}, \& {Blitz}}]{williams94}
{Williams}, J.~P., {de Geus}, E.~J., \& {Blitz}, L. 1994, \apj, 428, 693

\bibitem[{{Williams} {et~al.}(2022){Williams}, {Kreckel}, {Belfiore}, {Groves}, {Sandstrom}, {Santoro}, {Blanc}, {Bigiel}, {Boquien}, {Chevance}, {Congiu}, {Emsellem}, {Glover}, {Grasha}, {Klessen}, {Koch}, {Kruijssen}, {Leroy}, {Liu}, {Meidt}, {Pan}, {Querejeta}, {Rosolowsky}, {Saito}, {S{\'a}nchez-Bl{\'a}zquez}, {Schinnerer}, {Schruba}, \& {Watkins}}]{williams22}
{Williams}, T.~G., {Kreckel}, K., {Belfiore}, F., {et~al.} 2022, \mnras, 509, 1303

\bibitem[{{Wood} {et~al.}(2010){Wood}, {Hill}, {Joung}, {Mac Low}, {Benjamin}, {Haffner}, {Reynolds}, \& {Madsen}}]{Wood2010}
{Wood}, K., {Hill}, A.~S., {Joung}, M.~R., {et~al.} 2010, \apj, 721, 1397

\bibitem[{{Yu} {et~al.}(2021){Yu}, {Ho}, \& {Wang}}]{yu21}
{Yu}, S.-Y., {Ho}, L.~C., \& {Wang}, J. 2021, \apj, 917, 88

\bibitem[{{Zabel} {et~al.}(2020){Zabel}, {Davis}, {Sarzi}, {Nedelchev}, {Chevance}, {Kruijssen}, {Iodice}, {Baes}, {Bendo}, {Corsini}, {De Looze}, {de Zeeuw}, {Gadotti}, {Grossi}, {Peletier}, {Pinna}, {Serra}, {van de Voort}, {Venhola}, {Viaene}, \& {Vlahakis}}]{zabel20}
{Zabel}, N., {Davis}, T.~A., {Sarzi}, M., {et~al.} 2020, \mnras, 496, 2155

\end{thebibliography}

\begin{appendix}

\begin{table*}[!htbp]
\section{Best-fitting parameters}
\caption{Best-fitting parameters obtained from performing our statistical analysis to 22 spiral galaxies.}
\label{table1}
\centering
\renewcommand{\arraystretch}{1.25}
\begin{tabular}{ c  c  c  c  c  c  c  c  c c}
\hline\hline
Galaxy & Region & $t_{\rm CO}[{\rm Myr}]$ & $t_{\rm fb}[{\rm Myr}]$ & $\lambda[{\rm pc}]$ & $\epsilon_{\rm SF}[\%]$ & $\Sigma_{\rm CO,comp}$ & $\Sigma_{\rm SFR}$ & $\varepsilon_{\rm gas}$ & $t_{\rm ref}[Myr]$\\
\hline
IC 1954 & inter-arm & $9.5^{+4.1}_{-2.3}$ & $2.7^{+1.7}_{-1.3}$ & $233^{+108}_{-52}$ & $2.2^{+2.1}_{-1.0}$ & $2.3^{+1.2}_{-1.2}$ & $4.0^{+0.8}_{-0.8}$ & $6.7^{+3.8}_{-2.7}$ & $4.49^{+0.07}_{-0.20}$\\
 & spiral arm (*)& $22.4^{+13.7}_{-4.1}$ & $4.5^{+4.5}_{-1.6}$ & $163^{+86}_{-38}$ & $4.7^{+6.4}_{-1.9}$ & $4.5^{+2.2}_{-2.2}$ & $6.2^{+1.2}_{-1.2}$ & $4.1^{+1.3}_{-1.1}$ & $4.46^{+0.08}_{-0.20}$\\
\hline
NGC 0628 & inter-arm & $22.0^{+4.1}_{-4.4}$ & $2.8^{+0.8}_{-0.9}$ & $111^{+19}_{-16}$ & $6.8^{+5.3}_{-3.0}$ & $2.8^{+1.4}_{-1.4}$ & $6.0^{+1.2}_{-1.2}$ & $12.9^{+3.6}_{-3.0}$ & $4.47^{+0.08}_{-0.20}$\\
 & spiral arm & $27.1^{+3.3}_{-2.8}$ & $3.7^{+0.8}_{-0.6}$ & $90^{+15}_{-11}$ & $6.2^{+4.6}_{-2.6}$ & $6.7^{+3.4}_{-3.4}$ & $10.1^{+2.0}_{-2.0}$ & $7.3^{+1.1}_{-1.0}$ & $4.47^{+0.08}_{-0.20}$\\
\hline
NGC 1097 & inter-arm & $16.3^{+3.6}_{-2.6}$ & $1.0^{+0.7}_{-0.6}$ & $237^{+59}_{-31}$ & $3.4^{+2.8}_{-1.5}$ & $3.6^{+1.8}_{-1.8}$ & $5.0^{+1.0}_{-1.0}$ & $6.4^{+1.5}_{-1.3}$ & $4.35^{+0.09}_{-0.22}$\\
 & spiral arm (*)& $13.3^{+5.9}_{-3.0}$ & $1.3^{+1.2}_{-0.7}$ & $187^{+43}_{-20}$ & $2.2^{+2.1}_{-1.0}$ & $5.1^{+2.6}_{-2.6}$ & $5.6^{+1.1}_{-1.1}$ & $4.3^{+0.5}_{-0.5}$ & $4.37^{+0.09}_{-0.22}$\\
\hline
NGC 1300 & inter-arm & $14.4^{+3.5}_{-1.8}$ & $2.3^{+0.7}_{-0.4}$ & $386^{+84}_{-63}$ & $3.5^{+2.7}_{-1.5}$ & $1.3^{+0.6}_{-0.6}$ & $1.8^{+0.4}_{-0.4}$ & $18.1^{+8.4}_{-6.3}$ & $4.44^{+0.08}_{-0.21}$\\
 & spiral arm & $15.5^{+3.2}_{-3.6}$ & $4.5^{+1.1}_{-1.5}$ & $255^{+52}_{-48}$ & $2.1^{+1.7}_{-0.9}$ & $4.3^{+2.1}_{-2.1}$ & $4.0^{+0.8}_{-0.8}$ & $7.7^{+2.6}_{-2.1}$ & $4.45^{+0.08}_{-0.20}$\\
\hline
NGC 1365 & inter-arm & $32.1^{+16.7}_{-7.9}$ & $4.4^{+2.4}_{-1.5}$ & $360^{+125}_{-105}$ & $6.8^{+7.7}_{-3.0}$ & $4.7^{+2.3}_{-2.3}$ & $6.6^{+1.3}_{-1.3}$ & $5.7^{+2.7}_{-2.0}$ & $4.42^{+0.08}_{-0.21}$\\
 & spiral arm & $11.0^{+9.7}_{-2.4}$ & $3.8^{+4.5}_{-1.5}$ & $266^{+209}_{-77}$ & $2.1^{+3.0}_{-0.9}$ & $10.0^{+5.0}_{-5.0}$ & $12.3^{+2.5}_{-2.5}$ & $4.1^{+2.0}_{-1.4}$ & $4.44^{+0.08}_{-0.20}$\\
\hline
NGC 1385 & inter-arm (*) & $18.0^{+8.5}_{-4.3}$ & $3.2^{+2.4}_{-1.2}$ & $174^{+80}_{-35}$ & $7.5^{+7.8}_{-3.2}$ & $7.1^{+3.5}_{-3.5}$ & $19.4^{+3.9}_{-3.9}$ & $7.4^{+3.2}_{-2.3}$ & $4.55^{+0.07}_{-0.18}$ \\
 & spiral arm (*)& $10.0^{+3.7}_{-2.3}$ & $2.4^{+1.6}_{-1.0}$ & $176^{+99}_{-37}$ & $2.6^{+2.4}_{-1.1}$ & $7.3^{+3.7}_{-3.7}$ & $12.0^{+2.4}_{-2.4}$ & $7.3^{+3.3}_{-2.4}$ & $4.55^{+0.07}_{-0.18}$\\
\hline
NGC 1512 & inter-arm & $8.7^{+2.7}_{-1.4}$ & $1.8^{+0.4}_{-0.6}$ & $323^{+95}_{-99}$ & $3.1^{+2.8}_{-1.3}$ & $0.6^{+0.3}_{-0.3}$ & $1.5^{+0.3}_{-0.3}$ & $25.9^{+22.4}_{-13.6}$ & $4.42^{+0.08}_{-0.21}$\\
 & spiral arm & $13.9^{+2.5}_{-2.3}$ & $1.9^{+0.5}_{-0.5}$ & $245^{+29}_{-25}$ & $3.0^{+2.2}_{-1.4}$ & $1.5^{+0.8}_{-0.8}$ & $2.0^{+0.4}_{-0.4}$ & $11.2^{+2.1}_{-1.9}$ & $4.42^{+0.08}_{-0.21}$\\
\hline
NGC 1672 & inter-arm & $19.7^{+5.7}_{-4.7}$ & $3.8^{+1.9}_{-1.5}$ & $324^{+173}_{-83}$ & $6.1^{+5.1}_{-2.8}$ & $3.9^{+2.0}_{-2.0}$ & $8.2^{+1.6}_{-1.6}$ & $7.2^{+4.2}_{-2.8}$ & $4.43^{+0.08}_{-0.21}$\\
 & spiral arm (*)& $37.8^{+103.5}_{-7.7}$ & $9.2^{+29.1}_{-3.5}$ & $292^{+117}_{-80}$ & $7.1^{+20.4}_{-2.8}$ & $11.1^{+5.6}_{-5.6}$ & $12.9^{+2.6}_{-2.6}$ & $4.7^{+1.6}_{-1.3}$ & $4.44^{+0.08}_{-0.21}$\\
\hline
NGC 2090 & inter-arm & $7.8^{+1.7}_{-1.4}$ & $0.8^{+0.5}_{-0.4}$ & $235^{+69}_{-36}$ & $2.8^{+2.0}_{-1.3}$ & $1.1^{+0.6}_{-0.6}$ & $2.5^{+0.5}_{-0.5}$ & $12.8^{+5.9}_{-4.4}$ & $4.43^{+0.08}_{-0.21}$\\
 & spiral arm (*)& $11.5^{+8.6}_{-2.6}$ & $1.9^{+3.2}_{-1.0}$ & $132^{+47}_{-24}$ & $1.8^{+2.3}_{-0.7}$ & $5.8^{+2.9}_{-2.9}$ & $5.7^{+1.1}_{-1.1}$ & $3.1^{+0.9}_{-0.7}$ & $4.39^{+0.08}_{-0.21}$\\
\hline
NGC 2283 & inter-arm & $9.1^{+1.6}_{-1.8}$ & $2.7^{+0.5}_{-0.8}$ & $328^{+102}_{-74}$ & $3.9^{+2.8}_{-1.7}$ & $1.7^{+0.9}_{-0.9}$ & $5.2^{+1.0}_{-1.0}$ & $14.1^{+9.1}_{-6.2}$ & $4.44^{+0.08}_{-0.20}$\\
 & spiral arm (*)& $12.2^{+7.8}_{-3.5}$ & $3.8^{+4.1}_{-1.8}$ & $152^{+156}_{-40}$ & $5.4^{+6.4}_{-2.4}$ & $4.6^{+2.3}_{-2.3}$ & $13.2^{+2.6}_{-2.6}$ & $6.1^{+6.3}_{-3.2}$ &  $4.42^{+0.08}_{-0.22}$\\
\hline
NGC 2835 & inter-arm & $6.6^{+1.4}_{-1.1}$ & $0.9^{+0.5}_{-0.4}$ & $167^{+58}_{-32}$ & $2.0^{+1.6}_{-0.9}$ & $2.4^{+1.2}_{-1.2}$ & $4.9^{+1.0}_{-1.0}$ & $14.9^{+7.7}_{-5.6}$ & $4.59^{+0.06}_{-0.17}$\\
 & spiral arm & $10.7^{+2.6}_{-2.2}$ & $1.9^{+0.9}_{-0.7}$ & $132^{+39}_{-18}$ & $2.9^{+2.4}_{-1.3}$ & $4.6^{+2.3}_{-2.3}$ & $7.8^{+1.6}_{-1.6}$ & $8.8^{+3.6}_{-2.7}$ & $4.59^{+0.06}_{-0.21}$\\
\hline
NGC 2997 & inter-arm & $15.9^{+1.8}_{-1.9}$ & $3.4^{+0.6}_{-0.7}$ & $266^{+49}_{-37}$ & $4.0^{+2.9}_{-1.7}$ & $3.7^{+1.9}_{-1.9}$ & $6.5^{+1.3}_{-1.3}$ & $10.2^{+3.0}_{-2.5}$ & $4.43^{+0.08}_{-0.21}$\\
 & spiral arm & $18.7^{+6.2}_{-2.8}$ & $5.3^{+2.5}_{-1.2}$ & $215^{+64}_{-37}$ & $2.9^{+2.7}_{-1.2}$ & $14.0^{+7.0}_{-7.0}$ & $13.7^{+2.7}_{-2.7}$ & $5.4^{+1.4}_{-1.2}$ & $4.42^{+0.08}_{-0.17}$\\
\hline
NGC 3507 & inter-arm & $19.3^{+5.8}_{-4.0}$ & $2.1^{+1.5}_{-0.7}$ & $321^{+156}_{-59}$ & $4.5^{+3.7}_{-2.0}$ & $2.0^{+1.0}_{-1.0}$ & $2.9^{+0.6}_{-0.6}$ & $8.4^{+5.7}_{-3.7}$ & $4.45^{+0.08}_{-0.20}$\\
 & spiral arm (*)& $9.1^{+1.9}_{-1.3}$ & $2.6^{+1.1}_{-0.7}$ & $242^{+87}_{-42}$ & $1.4^{+1.1}_{-0.6}$ & $5.1^{+2.6}_{-2.6}$ & $4.8^{+1.0}_{-1.0}$ & $5.2^{+1.5}_{-1.2}$ & $4.44^{+0.08}_{-0.21}$\\
\hline
NGC 3627 & inter-arm & $14.2^{+3.0}_{-2.2}$ & $1.3^{+0.7}_{-0.5}$ & $224^{+37}_{-25}$ & $5.1^{+4.0}_{-2.2}$ & $5.8^{+2.9}_{-2.9}$ & $13.5^{+2.7}_{-2.7}$ & $7.7^{+1.3}_{-1.2}$ & $4.43^{+0.08}_{-0.21}$\\
 & spiral arm & $17.2^{+5.6}_{-3.2}$ & $3.1^{+1.8}_{-1.2}$ & $187^{+77}_{-40}$ & $3.1^{+2.7}_{-1.4}$ & $26.8^{+13.4}_{-13.4}$ & $31.2^{+6.2}_{-6.2}$ & $4.1^{+1.4}_{-1.1}$ & $4.44^{+0.08}_{-0.21}$\\
\hline
NGC 4254 & inter-arm & $17.0^{+2.6}_{-2.0}$ & $4.6^{+0.9}_{-1.0}$ & $264^{+74}_{-71}$ & $3.7^{+2.7}_{-1.6}$ & $8.4^{+4.2}_{-4.2}$ & $12.1^{+2.4}_{-2.4}$ & $7.1^{+3.1}_{-2.4}$ & $4.46^{+0.08}_{-0.20}$\\
 & spiral arm & $17.7^{+7.6}_{-2.2}$ & $5.0^{+3.1}_{-1.2}$ & $197^{+62}_{-39}$ & $3.7^{+3.5}_{-1.7}$ & $16.2^{+8.1}_{-8.1}$ & $20.6^{+4.1}_{-4.1}$ & $4.6^{+1.1}_{-0.9}$ & $4.45^{+0.08}_{-0.21}$\\
\hline
NGC 4303 & inter-arm & $31.2^{+11.8}_{-6.4}$ & $6.6^{+2.9}_{-1.9}$ & $320^{+77}_{-49}$ & $11.0^{+17.4}_{-4.9}$ & $6.7^{+3.4}_{-3.4}$ & $15.1^{+3.0}_{-3.0}$ & $6.0^{+2.1}_{-1.7}$ & $4.41^{+0.08}_{-0.21}$\\
 & spiral arm (*)& $17.4^{+5.0}_{-2.9}$ & $3.6^{+2.1}_{-1.4}$ & $233^{+96}_{-49}$ & $5.2^{+4.4}_{-2.2}$ & $17.3^{+8.6}_{-8.6}$ & $33.2^{+6.6}_{-6.6}$ & $4.2^{+1.4}_{-1.1}$ & $4.39^{+0.08}_{-0.20}$\\
\hline
NGC 4321 & inter-arm & $29.9^{+4.9}_{-3.7}$ & $3.8^{+1.2}_{-1.0}$ & $265^{+51}_{-38}$ & $7.9^{+5.9}_{-3.4}$ & $4.0^{+2.0}_{-2.0}$ & $6.8^{+1.4}_{-1.4}$ & $6.3^{+1.5}_{-1.3}$ & $4.41^{+0.08}_{-0.21}$\\
 & spiral arm (*)& $15.0^{+3.2}_{-2.0}$ & $2.6^{+1.1}_{-0.6}$ & $196^{+36}_{-24}$ & $2.4^{+1.9}_{-1.0}$ & $10.6^{+5.3}_{-5.3}$ & $11.1^{+2.2}_{-2.2}$ & $4.6^{+0.7}_{-0.6}$ &  $4.39^{+0.08}_{-0.22}$ \\
\hline
NGC 4535 & inter-arm & $33.2^{+73.0}_{-5.5}$ & $5.0^{+13.8}_{-1.1}$ & $267^{+56}_{-39}$ & $8.9^{+17.1}_{-3.8}$ & $2.4^{+1.2}_{-1.2}$ & $4.4^{+0.9}_{-0.9}$ & $8.9^{+2.4}_{-2.0}$ & $4.43^{+0.08}_{-0.21}$\\
 & spiral arm & $17.6^{+86.9}_{-3.0}$ & $4.6^{+29.0}_{-1.2}$ & $223^{+101}_{-51}$ & $2.2^{+10.7}_{-0.9}$ & $9.5^{+4.7}_{-4.7}$ & $7.7^{+1.5}_{-1.5}$ & $5.6^{+2.2}_{-1.7}$ & $4.42^{+0.08}_{-0.22}$ \\
\hline
NGC 4548 & inter-arm & $20.2^{+5.4}_{-4.7}$ & $2.4^{+0.9}_{-1.1}$ & $290^{+91}_{-66}$ & $3.3^{+2.7}_{-1.5}$ & $1.2^{+0.6}_{-0.6}$ & $1.3^{+0.3}_{-0.3}$ & $9.0^{+4.5}_{-3.3}$ & $4.39^{+0.08}_{-0.22}$\\
 & spiral arm (*)& $10.5^{+6.3}_{-2.4}$ & $1.4^{+1.9}_{-0.6}$ & $178^{+38}_{-20}$ & $1.0^{+1.2}_{-0.4}$ & $4.7^{+2.3}_{-2.3}$ & $2.8^{+0.6}_{-0.6}$ & $4.7^{+0.6}_{-0.5}$ & $4.40^{+0.08}_{-0.21}$\\
\hline
NGC 5248 & inter-arm & $12.9^{+3.9}_{-2.6}$ & $1.8^{+1.3}_{-0.8}$ & $177^{+39}_{-23}$ & $3.4^{+2.9}_{-1.4}$ & $4.2^{+2.1}_{-2.1}$ & $7.1^{+1.4}_{-1.4}$ & $9.2^{+2.1}_{-1.8}$ & $4.39^{+0.08}_{-0.21}$\\
 & spiral arm & $21.8^{+25.6}_{-5.4}$ & $4.2^{+4.8}_{-1.6}$ & $171^{+61}_{-42}$ & $3.4^{+5.0}_{-1.5}$ & $17.8^{+8.9}_{-8.9}$ & $18.5^{+3.7}_{-3.7}$ & $4.9^{+1.1}_{-0.9}$ & $4.39^{+0.08}_{-0.19}$\\
\hline
NGC 5643 & inter-arm & $17.5^{+15.1}_{-4.1}$ & $2.4^{+3.5}_{-1.1}$ & $194^{+73}_{-31}$ & $3.8^{+5.0}_{-1.6}$ & $6.8^{+3.4}_{-3.4}$ & $9.6^{+1.9}_{-1.9}$ & $7.8^{+4.3}_{-3.0}$ & $4.46^{+0.08}_{-0.20}$\\
 & spiral arm & $18.4^{+3.4}_{-2.7}$ & $3.4^{+0.9}_{-0.8}$ & $180^{+30}_{-22}$ & $4.0^{+3.1}_{-1.7}$ & $8.0^{+4.0}_{-4.0}$ & $11.1^{+2.2}_{-2.2}$ & $6.7^{+1.7}_{-1.5}$ & $4.47^{+0.08}_{-0.22}$\\
\hline
NGC 6744 & inter-arm & $24.9^{+4.5}_{-2.6}$ & $3.1^{+0.8}_{-0.5}$ & $157^{+18}_{-13}$ & $3.8^{+3.0}_{-1.6}$ & $3.3^{+1.6}_{-1.6}$ & $3.3^{+0.7}_{-0.7}$ & $8.7^{+1.6}_{-1.4}$ & $4.33^{+0.09}_{-0.23}$\\
 & spiral arm & $42.8^{+9.2}_{-4.7}$ & $5.3^{+1.6}_{-0.9}$ & $152^{+21}_{-14}$ & $3.1^{+1.6}_{-1.6}$ & $4.8^{+2.4}_{-2.4}$ & $3.8^{+0.8}_{-0.8}$ & $6.5^{+1.1}_{-1.0}$ & $4.30^{+0.09}_{-0.20}$\\
\hline
\end{tabular}
\tablefoot{(*) indicates the environments in which the ratio $\lambda/l_{\rm ap,min}<1.5$, implying that the independent star-forming regions are not sufficiently resolved. In these cases the estimates of $t_{\rm fb}$ and $\lambda$ are only 1-$\sigma$ upper limits, while the estimate of $t_{\rm CO}$ is unaffected by resolution issues (see section 4.3.6 in \citealt{kruijssen18}). There is a strong bias for this issue to occur preferentially in the spiral arms as the surface density and the number of regions is higher in this environment.}
\end{table*}
\end{appendix}

\end{document}